\documentclass[aps,prc,twocolumn,superscriptaddress,amsfonts,amsmath,amssymb,showpacs,floatfix]{revtex4-1}
\usepackage[pdftex]{graphicx} 
\usepackage{hyperref,longtable}
\usepackage{color}
\usepackage[dvipsnames,svgnames,x11names,hyperref]{xcolor}
\hypersetup{colorlinks,citecolor=[RGB]{132 112 255},filecolor=[RGB]{220 20 60},linkcolor=[RGB]{191 62 255},urlcolor=[RGB]{138 43 226}}
\usepackage{float}
\usepackage{footnote}

\begin{document}

\title{An experimental study of the rearrangements of valence protons and neutrons amongst single-particle orbits during double {\boldmath$\beta$} decay in $^{100}$Mo.}

\author{S.~J.~Freeman}
\email[Correspondence to: ]{sean.freeman@manchester.ac.uk}
\affiliation{School of Physics and Astronomy, The University of Manchester, Manchester, M13 9PL, United Kingdom}
\author{D.~K.~Sharp}
\affiliation{School of Physics and Astronomy, The University of Manchester, Manchester, M13 9PL, United Kingdom}
\author{S.~A.~McAllister}
\affiliation{School of Physics and Astronomy, The University of Manchester, Manchester, M13 9PL, United Kingdom}
\author{B.~P.~Kay}
\altaffiliation[Current address: ]{Argonne National Laboratory, Physics Division, Argonne, Illinois 60439, USA}
\affiliation{Department of Physics, University of York, York, YO10 5DD, United Kingdom}
\author{C.~M.~Deibel}
\altaffiliation[Current address: ]{Department of Physics and Astronomy, Louisiana State University, Baton Rouge, LA 70803,
USA.}
\affiliation{Argonne National Laboratory, Physics Division, Argonne, Illinois 60439, USA}
\affiliation{Joint Institute for Nuclear Astrophysics, Michigan State University, East Lansing, Michigan 48824, USA}
\author{T.~Faestermann}
\affiliation{Physik Department E12, Technische Universit\"at M\"unchen, D-85748 Garching, Germany}
\affiliation{Maier-Leibnitz Laboratorium der M\"unchner Universit\"aten (MML), D-85748 Garching, Germany}
\author{R.~Hertenberger}
\affiliation{Maier-Leibnitz Laboratorium der M\"unchner Universit\"aten (MML), D-85748 Garching, Germany}
\affiliation{Fakult\"at f\"ur Physik, Ludwig-Maximillians Universit\"at M\"unchen, D-85748 Garching, Germany}
\author{A.~J.~Mitchell}
\altaffiliation[Current address: ]{Department of Nuclear Physics, Research School of Physics and Engineering, The Australian National University, Canberra, ACT 2601, Australia.}
\affiliation{School of Physics and Astronomy, The University of Manchester, Manchester, M13 9PL, United Kingdom}
\author{J.~P.~Schiffer}
\affiliation{Argonne National Laboratory, Physics Division, Argonne, Illinois 60439, USA}
\author{S.~V.~Szwec}
\affiliation{School of Physics and Astronomy, The University of Manchester, Manchester, M13 9PL, United Kingdom}
\author{J.~S.~Thomas}
\affiliation{School of Physics and Astronomy, The University of Manchester, Manchester, M13 9PL, United Kingdom}
\author{H.-F.~Wirth}
\affiliation{Maier-Leibnitz Laboratorium der M\"unchner Universit\"aten (MML), D-85748 Garching, Germany}
\affiliation{Fakult\"at f\"ur Physik, Ludwig-Maximillians Universit\"at M\"unchen, D-85748 Garching, Germany}

\date{\today}

\begin{abstract}
The rearrangements of protons and neutrons amongst the valence single-particle orbitals during  double $\beta$ decay of $^{100}$Mo have been determined by measuring cross sections in  ($d$,$p$), ($p$,$d$), ($^3$He,$\alpha$) and ($^3$He,$d$) reactions on  $^{98,100}$Mo and $^{100,102}$Ru targets. The  deduced  nucleon occupancies  reveal significant discrepancies when compared with theoretical calculations; the same calculations have previously been used to determine the nuclear matrix element associated with the decay probability of  double $\beta$ decay of the $^{100}$Mo system.

\end{abstract}


\maketitle
\section{Introduction}

Over the past decade  observations of neutrino flavor oscillations have provided fundamental information about the relative masses of neutrinos and mixing angles. However, the process of neutrinoless double $\beta$ ($0\nu2\beta$) decay, if it is ever observed, would establish that the neutrino is a  Majorana fermion and could be a way of obtaining the absolute scale for neutrino mass eigenstates. During such a decay, two neutrons in the ground state of an even-even nucleus transform into protons, usually in the ground-state of the final nucleus, with the simultaneous emission of two electrons. The rate of decay can be expressed as a product of three independent factors (see for example, Ref.~\cite{Vogel}):

$${1\over T^{0\nu}_{1/2}}=G_{0\nu}\vert M_{0\nu} \vert^2  \langle m_{\beta\beta}\rangle^2.$$

Here,  $G_{0\nu}$ is the so-called phase-space factor and the information on the absolute mass scale appears via the term $\langle m_{\beta\beta}\rangle$, which is the effective neutrino mass in the simplest theoretical decay mechanisms. The dependence on nuclear structure is held in the nuclear matrix element $ M_{0\nu}$ that encapsulates the connection between initial and final nuclear states.

Both the nuclear matrix element and phase space factor are required if the absolute neutrino mass scale is to be deduced from a future half-life measurement of  neutrinoless double $\beta$ decay. Indeed, estimates of these quantities are also critical in planning projects that set out to search for the decay process; the extremely low expected decay probabilities corresponding to $T^{0\nu}_{1/2}\gtrsim10^{25}$~yr require extremely large-scale, low-background source-detector systems. 

Methods used in the calculation of phase-space factors are relatively well refined (see for example, Ref.~\cite{Kotila13} and references therein). However, there are significant difficulties associated with obtaining values of the nuclear matrix elements. Firstly, there are no other nuclear processes that directly probe the same matrix element, besides $0\nu2\beta$ decay itself. Secondly, even in a future era where $0\nu2\beta$ decay may have been unambiguously observed, it is unlikely that systematic phenomenological methods, which are common approaches to developing an understanding of many other complex nuclear characteristics, will be able to be applied in this case. The scale of investment required in attempts to observe a process with such low expected decay probabilities is such that, even if $0\nu2\beta$ decay is eventually observed, we are unlikely to have data on more than one or two isotopes for a considerable period of time, making phenomenology difficult. Therefore, in order to proceed, robust theoretical calculations of the nuclear matrix elements must be developed.

There has been significant progress in the understanding of the theoretical calculation of nuclear matrix elements for $0\nu2\beta$ decay over the last decade. As an illustration, in 2004, a provocative article \cite{Bahcall} suggested that the variation in the size of matrix elements calculated in different ways could be as much as two orders of magnitude. Whilst more recent developments have reduced the variation somewhat (see for example Ref.~\cite{Vogel}),  the convergence of different theoretical approaches in itself is no guarantee that they are correct.  It is also important to bear in mind that the matrix element appears as a square in the decay probability, increasing the variation between different models in their predictions of observable quantities.

One way forward is to determine which accessible properties of nuclei  are most directly relevant to the matrix elements. These properties can then be measured and the results used to gauge to what extent they can be reproduced by the models used to calculate the matrix elements. In this way, the calculational frameworks adopted can be constrained by comparison with other pertinent nuclear observables.

Double $\beta$ decay involves the decay of two neutrons into two protons within a nuclear system. The simplest of several mechanisms proposed involves a pair of virtual neutrinos  as the nucleons transform. If neutrinos are Majorana in nature, the annihilation of the virtual neutrinos leads to the neutrinoless version of the decay. The whole process is often viewed as proceeding via the excitation of states in the  intermediate isobaric nucleus between the parent and daughter (see, for example, \cite{Vogel} for more details). The short distance scales within the nuclear system imply the involvement of large virtual momenta (up to $\sim$100~MeV/c), leading to high virtual excitation energies (50-100~MeV) in the intermediate system and  angular momenta up to $\sim7-8\hbar$. Unlike the two-neutrino form of double beta decay, which proceeds via  a small number of virtual 1$^+$ states at relatively low excitation, neutrinoless double  $\beta$ decay involves a large number of states to high excitation. It therefore seems unlikely that neutrinoless double  $\beta$ decay exhibits strong sensitivity to the detailed structure of the intermediate nucleus. However, the ground states of the initial and final nuclei must play a role in determining the value of the matrix element. If there are significant rearrangements  of other nucleons, beyond the simple conversion of the two neutrons into  protons, the decay rates may be diminished; a change in nuclear deformation accompanying the decay is an extreme example of such a rearrangement. Such inhibition is common in other types of nuclear processes. The differences in the occupation of the valence single-particle states before and after the decay characterises such rearrangements, which are likely to have important consequences for the matrix element. 

Determining the valence populations of neutrons and protons, and the difference in these populations between initial and final states, addresses a critical ingredient of the overlap that determines the matrix elements \cite{Focus}. For example, we carried out systematic measurements of the valence proton and neutron occupancies in $^{76}$Ge and $^{76}$Se, a potential $0\nu2\beta$ parent-daughter system \cite{Ge_n, Ge_p}. Several authors have revisited theoretical predictions in the light of this data, leading to a reduction in the difference between predictions of the matrix elements  based on the quasi-particle random phase approximation  (QRPA) compared to those made using the interacting shell model (as examples, Refs.~\cite{Simkovic} and \cite {Menendez}). Several measurements have been made to characterise other systems and provide further benchmarks for theoretical approaches \cite{Te, Entwisle, Szwec}.
 
Here we report on a consistent set of single-nucleon transfer experiments that have been used to determine the valence nucleon occupations for $^{100}$Mo and $^{100}$Ru. This builds on our previous experimental study \cite{Thomas} of the validity of the BCS approximation in these nuclei, which is a basic assumption of the widely used QRPA method. Nucleon occupancies in $^{98}$Mo and $^{102}$Ru were also measured and used as consistency checks. $^{100}$Mo and $^{100}$Ru are parent and daughter for a potential $0\nu2\beta$ decay whose $Q$ value makes it a good candidate in which to observe the process \cite{Vogel}. There are several experiments that propose searches for double $\beta$ decay of $^{100}$Mo. For example, it is one of the isotopes that may be used in the SuperNEMO $0\nu2\beta$ decay experiment and was a source in the predecessor NEMO-3 \cite{Arnold1, Arnold2}. Other examples include the AMoRE  \cite{amore} and CUPID/LUMINEU \cite{lumineu} experiments which have proposed plans to use a cryogenic scintillation detector based on molybdate crystals.

This potential double-$\beta$-decay system lies toward the edge of an interesting region of the chart of nuclides. Nuclei with Z$\sim$40 exhibit a sudden onset of deformation resulting in a dramatic shape change from spherical to prolate shapes near $N=60$. The first indication of this transition came from early studies of $\gamma$ emission from spontaneous fission fragments \cite{johansson}, measurements that have since been refined significantly with the improvements in detection technology (see for example \cite{gavin}). For molybdenum and ruthenium isotopes, the evolution in shape persists, but is more gradual in nature. For example, a smoother shape transition has recently been inferred in mean-square charge radii of molybdenum fission fragments \cite{charlwood} from $A\sim 98$ to 104 using laser spectroscopy of separated singly-charged ions. Classical optical  spectroscopy of enriched isotopes of ruthenium \cite{King} paints a similar picture, although  there is some evidence for triaxial shapes in ruthenium fission fragments beyond $N=60$ \cite{shannon}. The transitional nature of $^{100}$Mo is also clear from  pair transfer studies. For example, our recent ($p$,$t$) reaction studies on targets of $^{98,100}$Mo and $^{100,102}$Ru \cite{Thomas} found that  95\% of the neutron pair transfer strength to 0$^+$ states is contained in the ground-state transition, except for the reaction leading to $^{98}$Mo, where a state at 735 keV was populated with $\sim$20\% of the ground-state transition strength. This transitional nature, and potential structural differences between parent and daughter, are likely to present challenges for the calculations of the associated $0\nu2\beta$ decay matrix elements.

Over many years, data has been accumulated on single-nucleon transfer data that might yield the occupancies required to constrain calculations of the $0\nu2\beta$ matrix elements. Molybdenum isotopes have been studied in neutron transfer experiments.  There are several published studies of the $(d,p)$  and $(p,d)$ reactions on $^{100}$Mo\cite{101Mo_dp1,101Mo_dp2,101Mo_dp3,101Mo_dp4,101Mo_dp5,99Mo_101Mo_dp,99Mo_pd1,99Mo_pd2, 97Mo_99Mo_pd1, 97Mo_99Mo_pd2}. Studies of both these reactions, albeit fewer in number, have been made on targets of $^{98}$Mo  \cite{99Mo_dp1,99Mo_dp2,99Mo_101Mo_dp,97Mo_pd} and $^{102}$Ru \cite{103Ru_dp1,103Ru_dp2,103Ru_dp3,103Ru_dp4,103Ru_pd}.  Neutron addition has been performed on $^{100}$Ru \cite{101Ru_dp1,101Ru_dp2}, although the neutron removal with the $(p,d)$ reaction has not. Of these four targets, only $^{100}$Mo has been studied in the ($^3$He,$\alpha$) reaction \cite{99Mo_ha}, which is required to provide good matching for large angular-momentum transfer. Where data does exist in the literature, the experiments were performed at different times using different experimental techniques, different bombarding energies, different ranges of excitation energy and so on. Reaction modeling has been employed differently in each case, using a variety of computer codes and employing a host of  different approximations and potential choices. In some cases, measured cross sections have not been published. As a result, the existing literature, whilst useful in establishing many spin-parity assignments to relevant states,  has neither the overall precision nor the consistency required to determine the changes in neutron occupancies between parent and daughter in this potential $0\nu2\beta$ system. 

With regards to proton transfer  reactions,  ($^3$He,d) studies have been made on $^{98,100}$Mo \cite{99Tc_hd1,99Tc_hd2,101Tc_hd}, but not on the relevant ruthenium isotopes.  The majority of previous studies were done at significantly worse resolution than the current work;  resolution was one of the contributory factors in determining how high in excitation energy measurements could be undertaken. The comments above concerning the consistency of experimental approach and reaction modeling are also pertinent for the proton transfer data in the literature.

In the current work, several transfer reactions have been employed. The $(p,d)$ and $(d,p)$ reactions were used to gain spectroscopic information on the low-$\ell$ valence neutron states. In these reactions, we have determined the normalization of the necessary reaction model calculations  by requiring the sums of strength for addition and removal to be equal to the total degeneracy of the relevant orbits. The ($^3$He,$\alpha$) reaction was used to measure high-$\ell$ states, with a reaction normalization determined by the requirement that the sum of associated high-$\ell$ strength and the normalized low-$\ell$ strength from the $(p,d)$  reactions yields the expected number of valence neutrons. Using these reaction normalizations, neutron occupancies are deduced  from the neutron-removing reactions for the $0g_{7/2}$, $1d$, $2s_{1/2}$ and $0h_{11/2}$ orbitals.

For protons,  the ($^3$He,$d$) reaction was used to determine  proton vacancies. This reaction is reasonably well matched for all the valence orbitals of interest and was therefore normalized by requiring the total extracted transfer strengths to sum to the total number of valence proton holes. Orbital vacancies were then deduced for the proton $0g_{9/2}$, $1p$ and $0f_{5/2}$  orbitals. 

This current publication is organised in the following way. Common aspects of the experimental methodology will be discussed first. The features of the neutron and proton transfer reaction experiments will be considered in separate sections covering specific features of the results and analysis, the spin-parities of the populated states and features of the transfer strength distributions. The approach used to normalize the reaction modeling for the transfer of both types of nucleon will be described followed by a discussion of the extracted occupancies and their uncertainties. The deduced proton and neutron occupancies will finally be compared to those used in theoretical calculations of the double $\beta$ decay matrix elements and some conclusions are reached. For the sake of brevity, the detailed experimental data is available as Supplemental Material \cite{Supplemental} and discussion here will concentrate on more global information such as summed strengths.

\section{Experimental Methods}

Beams of the required ions  were delivered by the MP tandem accelerator at the Maier-Leibnitz Laboratorium of the Ludwig-Maximilians Universit\"at and the Technische Universit\"at M\"unchen. They were used to bombard isotopically enriched targets of $^{100}$Mo (97.39\%), $^{100}$Ru (96.95\%), $^{98}$Mo (97.18\%) and $^{102}$Ru (99.38\%) with nominal thicknesses of 100~$\mu$g/cm$^2$,  which were evaporated onto thin carbon foils with thicknesses in the range of 8-20~$\mu$g/cm$^2$. Beam currents were measured using a Faraday cup behind the target ladder connected to a  current integrator and were typically between 500 and 700~nA.

Light reaction products were momentum analyzed using a Q3D magnetic spectrometer \cite{Q3D}. The spectrometer entrance aperture, which defines the solid-angle acceptance of the system, was set at  a nominal value of 13.9~msr throughout the entire experiment to minimize systematic uncertainties. At the focal plane of the spectrometer, a multi-wire gas proportional counter backed by a plastic scintillator was used to measure position, energy loss and residual energy of the ions passing through it \cite{Wirth}. The focal-plane position was determined by reading out 255 cathode pads, positioned every 3.5mm  across the counter. Each pad was equipped with an individual integrated preamplifier and shaper. Events were registered when three to seven adjacent pads had signals above threshold. The digitized signals on active pads were then fitted with a Gaussian line shape resulting in a position measurement with a resolution that was better than 0.1~mm. Outgoing particles were identified by a combination of their magnetic rigidity and their energy-loss characteristics in the proportional counter and scintillator.

In order to extract absolute cross sections, the product of the target thickness and the solid angle of the spectrometer entrance aperture was determined using Coulomb elastic  scattering. The data were collected in two distinct running periods and elastic scattering was performed separately for both. In the first run, elastic scattering of 12-MeV $^3$He ions  at $\theta_{\rm lab}$=25$^\circ$ was used and in the second, similar measurements with 9-MeV deuterons at $\theta_{\rm lab}$=12$^\circ$. The elastic scattering cross sections under these conditions are predicted to be within 2\% and 4\% of the Rutherford scattering formula, respectively, according to optical-model calculations performed with the potentials discussed below. Lower beam currents were used for the elastic-scattering measurements compared to the transfer reactions, requiring a different scale on the current integrator. The calibrations of all the scales used during the experiment were determined using a calibrated current source to ensure that relative values are well known. Consistent results were obtained from the two different running periods and the overall uncertainty in the cross sections deduced using this approach was estimated to be around  5\%.

Reaction modeling must be performed to extract spectroscopic strengths from the measured cross section and the associated calculations were performed using the distorted-wave Born approximation (DWBA). The approximations involved are best met at the first maximum of the angular distribution of transfer products. In order to extract robust spectroscopic factors, data were therefore taken at the angles corresponding to these maxima for the relevant $\ell$ transfers  in each reaction. Measurements  were also made at some other angles when time allowed. The angles where data were taken are summarised  in Table \ref{Ang-Table} for each reaction. Although much of the measured strength was associated with states having pre-existing spin-parity assignments, the resulting sets of data map out  angular distributions that were  sufficient to discriminate between different angular momentum transfers to confirm or, where necessary,  make $\ell$ assignments. The comparison between the differently matched reactions,  $(p,d)$ and ($^3$He,$\alpha$), helps to extend the range of momentum transfers investigated in the angular distributions and the differences in cross section assisted some of the $\ell$ assignments, as discussed below.

Given the large number of  cross section measurements made to states populated over a range of several MeV in excitation, in four different reactions at several angles and on four different targets, the state-by-state cross section data is given in the Supplemental Material \cite{Supplemental}.

\begin{figure}
\centering
\includegraphics[scale=0.4]{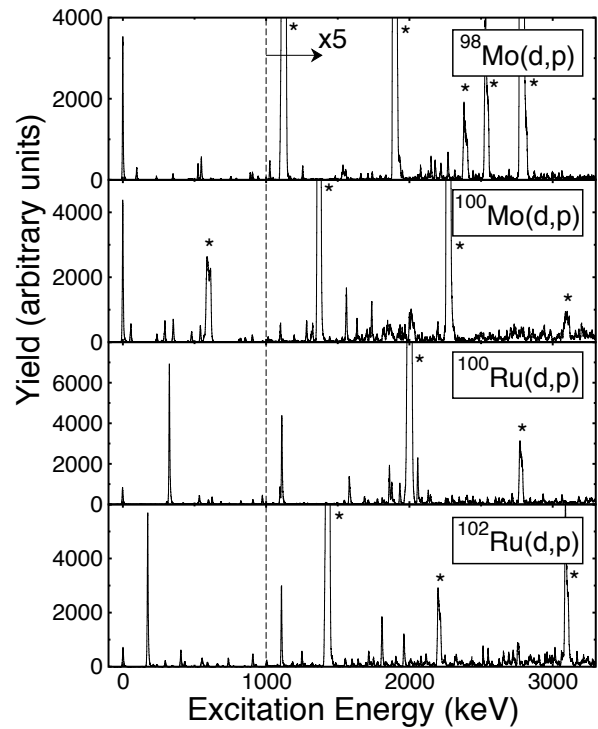}
\caption{\label{dp} Spectra of protons from the ($d$,$p$) reaction on targets of  $^{98}$Mo, $^{100}$Mo, $^{100}$Ru and $^{102}$Ru  at a laboratory angle of 8$^\circ$ as a function of the excitation energy in the residual nucleus. The portions of the spectra to the right of the dotted line have been scaled up by a factor of five. The broader peaks that appear in these spectra are reactions on light target contaminants, the strongest of which are marked by an asterisk.}
\end{figure}

\begin{figure}
\centering
\includegraphics[scale=0.4]{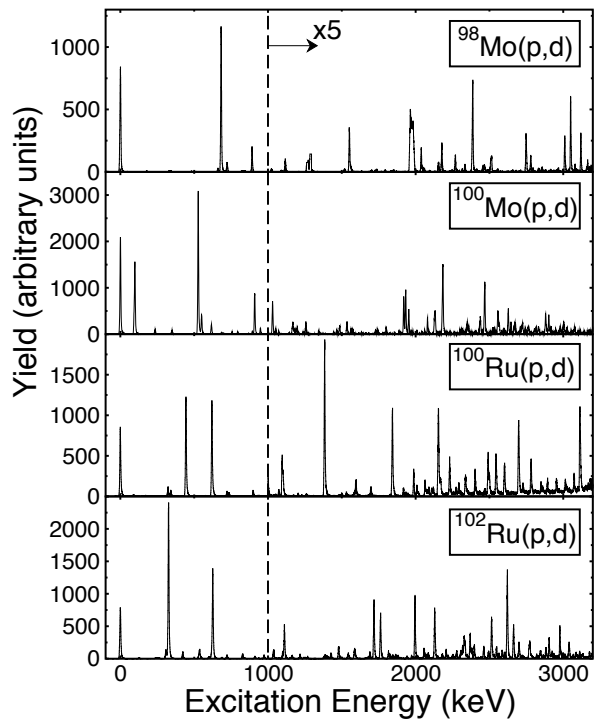}
\caption{\label{pd} Spectra of deuterons from the ($p$,$d$) reaction on targets of  $^{98}$Mo, $^{100}$Mo, $^{100}$Ru and $^{102}$Ru  at a laboratory angle of 6$^\circ$ as a function of the excitation energy in the residual nucleus. The portions of the spectra to the right of the dotted line have been scaled up by a factor of five.}
\end{figure}

\begin{figure}
\centering
\includegraphics[scale=0.4]{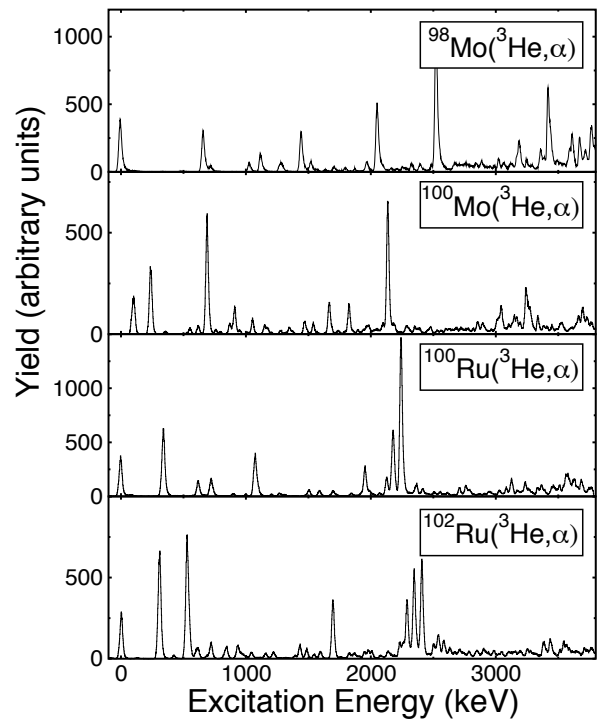}
\caption{\label{ha_sp} Spectra of $\alpha$ particles from the ($^3$He,$\alpha$) reaction on targets of  $^{98}$Mo, $^{100}$Mo, $^{100}$Ru and $^{102}$Ru  at a laboratory angle of 10$^\circ$ as a function of the excitation energy in the residual nucleus.  }
\end{figure}

\subsection{Neutron Transfer Reactions} 

The neutron-removal reactions, ($^3$He,$\alpha$) and $(p,d)$, were carried out with beams of $^3$He ions at an energy of 36~MeV and protons at 24~MeV, respectively.  The $(d,p)$ neutron-adding reaction was also performed using a deuteron beam at 15~MeV. Data were recorded up to excitation energies of at least 3~MeV in each residual nucleus. For the $(d,p)$ and $(p,d)$ reactions, this was achieved using three different magnet settings, arranged so that the  subsequent spectra overlapped in excitation by at least  100~keV.  The lower dispersion associated with the magnet settings for the  ($^3$He,$\alpha$) reaction enabled data to be recorded at one magnet setting. Figs.~\ref{dp}, \ref{pd} and \ref{ha_sp} show typical energy spectra of outgoing ions from these reactions. The spectra were calibrated using previously observed strongly populated final states \cite{A=97,A=99,A=101,A=103}. 

Excitation energies were estimated to be accurate to better than $\sim$3~keV for the $(d,p)$ reaction and around $\sim$2~keV for the $(p,d)$ reaction. For the ($^3$He,$\alpha$) reaction, low-lying states are accurate to $\sim$5~keV, rising to $\sim$10~keV at the higher excitation energies measured. Typical energy resolutions obtained were  $\sim$30~keV FWHM for ($^3$He,$\alpha$) and $\sim$8~keV FWHM for  $(p,d)$ and $(d,p)$ reactions. 

Peaks corresponding to reactions on carbon and oxygen target contaminants are present in the $(d,p)$ spectra with larger widths than those from the main target material due to their larger kinematic shift. These contaminant peaks obscured groups of interest at some angles, but the difference in their kinematic shifts meant that angles were always available where clean measurements could be made. The spectra were also checked carefully for the presence of any peaks arising from isotopic contaminants in the target material and these were excluded from subsequent analysis.

\begin{table}[H]
\caption{\label{Ang-Table} List of laboratory angles at which measurements were made for each of the reactions used. Due to target problems, data were not measured for the $^{98}$Mo($^3$He,$d$) reaction at 14$^\circ$ and 22$^\circ$.\\}
\begin{ruledtabular}
\begin{tabular}{l l }
Reaction & Laboratory Angles\\
\hline
\\
 $(p,d)$ & 6$^\circ$, 18$^\circ$, 31$^\circ$, 40$^\circ$\\
$(d,p)$ & 8$^\circ$, 18$^\circ$, 27$^\circ$, 33$^\circ$\\
($^3$He,$\alpha$)&10$^\circ$, 15$^\circ$, 20$^\circ$, 25$^\circ$\\
($^3$He,$d$) & 6$^\circ$, 10$^\circ$, 14$^\circ$, 18$^\circ$, 22$^\circ$\\\\
 \end{tabular}
 \end{ruledtabular}
 \end{table}


The differences in the kinematic matching between the two different neutron-removal reactions are apparent in the spectra.  For example, the $\ell=0$ ground state in $^{99}$Mo is clearly visible in the $(p,d)$ spectrum (Fig.~\ref{pd}) with a cross section of 2.98~mb/sr. However, it is hardly discernible at all in Fig.~\ref{ha_sp}, having a cross section of only 7~$\mu$b/sr in the ($^3$He,$\alpha$) reaction at 10$^\circ$, and approaches the observation limit of around 1~$\mu$b/sr at other angles. The ground state is only visible at all due to the low level density in this region; other excited $\ell=0$ transitions in the ($^3$He,$\alpha$) reaction are generally much weaker and obscured by stronger transitions.

\begin{figure}
\centering
\includegraphics[scale=0.5]{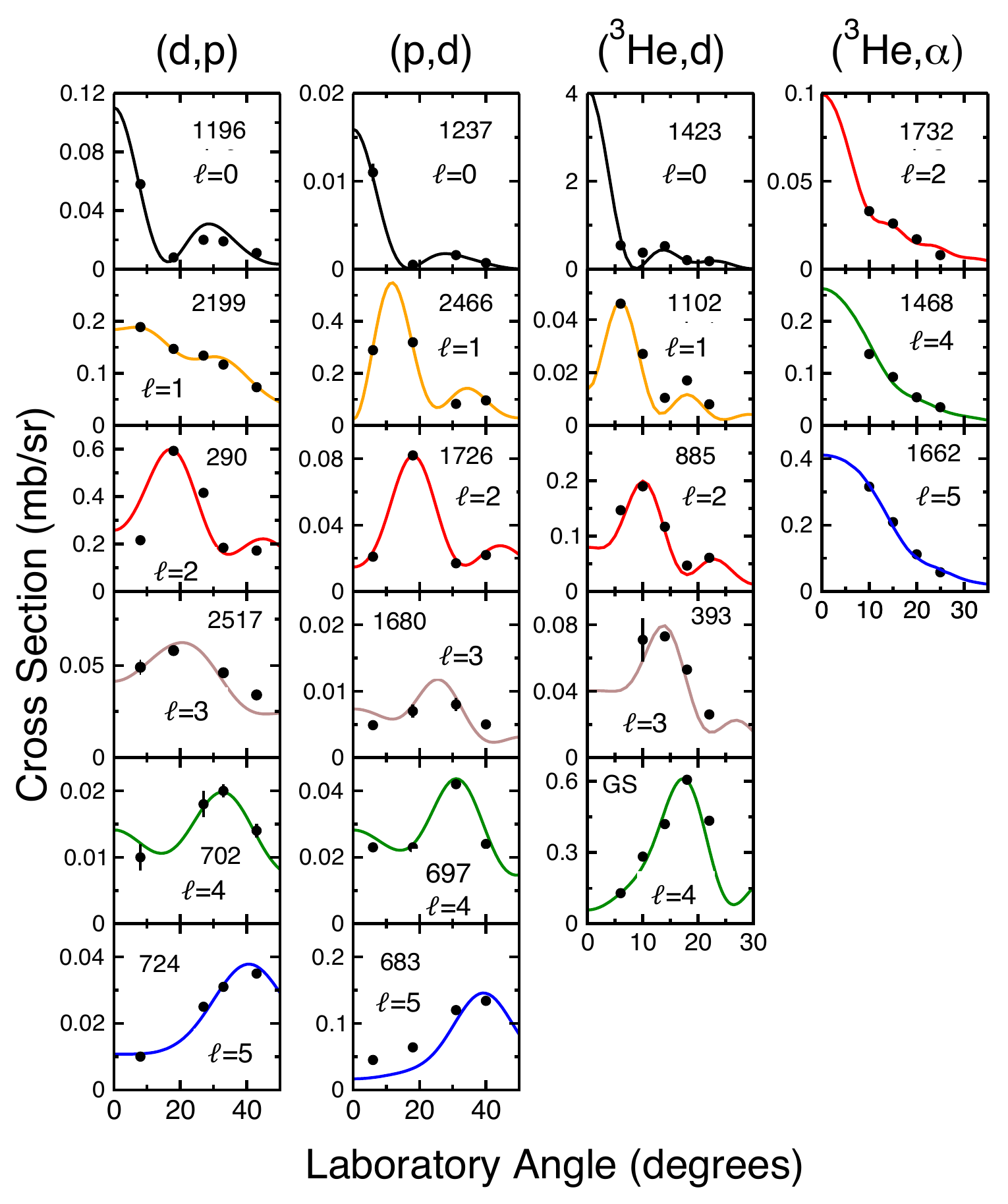}
\caption{\label{Dists} (Color online) Examples of angular distributions for the $(d,p)$, $(p,d)$, ($^3$He,d) and ($^3$He,$\alpha$) reactions on a $^{100}$Mo target. An example of each $\ell$ value is shown and compared to the results of DWBA calculations using parameters listed in Section III; $\ell=0$ (black),  $\ell=1$ (orange), $\ell=2$ (red), $\ell=3$ (brown), $\ell=4$ (green) and $\ell=5$ (blue).  Transitions with  $\ell=0$, 1 and 3 were not strongly observed in the  ($^3$He,$\alpha$) reaction. The angular distributions are labelled with $\ell$ value and the excitation energy in the residual system in units of keV.}
\end{figure}

\begin{figure*}
\centering
\includegraphics[scale=0.35]{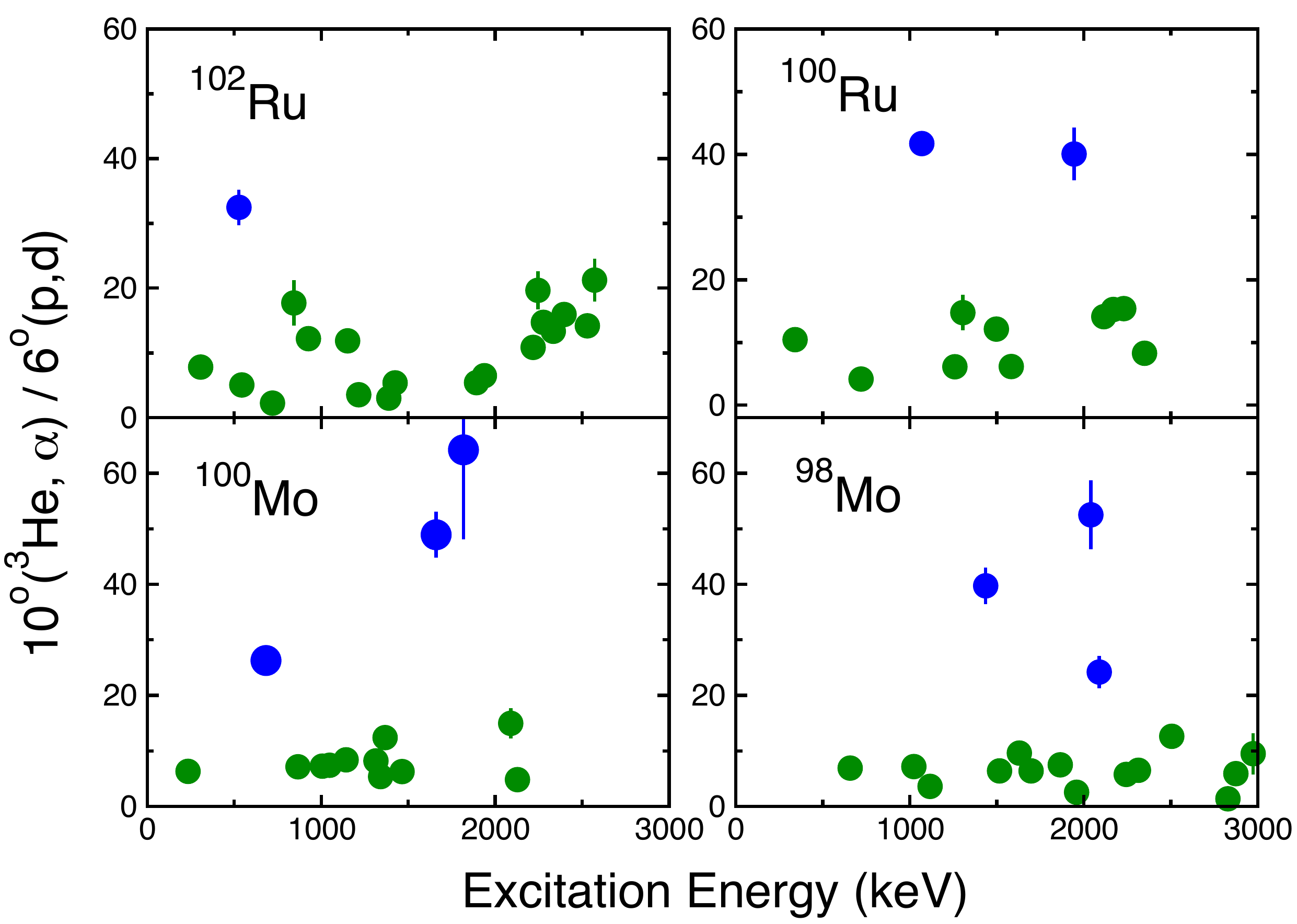}
\caption{\label{ratios} (Color online) The ratio of the cross section leading to the population of states in the ($^3$He,$\alpha$) reaction at a laboratory angle of 10$^{\circ}$ to that leading to the same states in the $(p,d)$ reaction at 6$^{\circ}$ for $\ell=4$ (green) and $\ell=5$ (blue) transitions, plotted as a function of excitation energy in the residual nucleus. The plots are labeled by the target isotope.}
\end{figure*}

For many of the states populated in the residual odd nuclei, angular-momentum quantum numbers have already been determined in a variety of previous studies that are summarized in Refs.~\cite{A=97,A=99,A=101,A=103}. Overall more than 85\% of the transfer strength used in the sum-rule analysis from which  the occupancies are extracted (as described below) is associated with states that have a previously determined  assignment. Where new assignments were made or previous assignments checked, this was done on the basis of the angular distribution of the light reaction product and a comparison of the cross section between the differently matched neutron-removal reactions. Some examples of angular distributions are shown in Fig. \ref{Dists} where the first maxima clearly appear at higher angles for higher $\ell$ transfers, except for the mismatched ($^3$He,$\alpha$) reaction where the forward-peaked shapes are less characteristic of the $\ell$ transfer. The strategy adopted when making new assignments was to use the shape of the distributions from $(p,d)$ and $(d,p)$ reactions, but confirm any high-$\ell$ assignments using the comparison of the cross sections from $(p,d)$ and ($^3$He,$\alpha$) reactions. Examples of the latter are shown in Fig. \ref{ratios} where the ratio of these cross sections at forward angles for $\ell=4$ and $\ell=5$ transitions is plotted. The momentum matching was such that $\ell=5$ transitions are characterised by larger ($^3$He,$\alpha$) to $(p,d)$ cross section ratios than those with $\ell=4$. Cross section ratios for transitions with $\ell<4$, not shown in Fig. \ref{ratios}, are smaller by factors of ten compared with those plotted. Whilst most of the consideration of such ratios was done using data at 6$^\circ$ for the $(p,d)$ reaction and 10$^\circ$ for the ($^3$He,$\alpha$)  reaction, ratios involving cross sections at other laboratory angles have similar features and were  used where needed, as noted in the Supplemental Material \cite{Supplemental}.  This assignment methodology produced results that were consistent with previous assignments where they are available in the literature.

Most of the states with significant contributions to the sum-rule analysis discussed below  have assignments from previous work. There are a few strong states where new assignments have been made here, most notably in the neutron-removal reactions on Mo targets populating states via $\ell=5$ transfer.  Newly assigned states at 2.043 and 2.089~MeV in $^{97}$Mo, carry 59\%  and  7\% of the measured $\ell=5$ strength respectively in that system. Similarly, two newly assigned states at 1.662 and 1.818~MeV in $^{99}$Mo contribute a third of the observed $\ell=5$ strength in $^{101}$Mo. These  states have $(p,d)$ cross sections that peak at the most backward angles studied and the ratios of  ($^3$He,$\alpha$) to $(p,d)$ cross sections are large and consistent with other $\ell=5$ transitions.  In the $(d,p)$ reaction, around a third of the $\ell=2$ strength on each of the molybdenum targets was from states with new assignments. In the $(p,d)$ reaction, the only significant newly assigned strength of relevance to the later analysis was the addition of  new $\ell=0$ strength in $^{99}$Ru. Much of this newly assigned low-$\ell$ strength arises from extending the excitation-energy range over which  measurements have been made; for example, states populated in $(d,p)$ reactions on molybdenum targets are only reported to around 1.5~MeV in the literature \cite{A=99,A=101}. For some weaker newly observed transitions, only tentative assignments were possible, but the contribution of these to the overall sum-rule analysis is naturally very small.

\begin{figure*}
\centering
\includegraphics[scale=0.6]{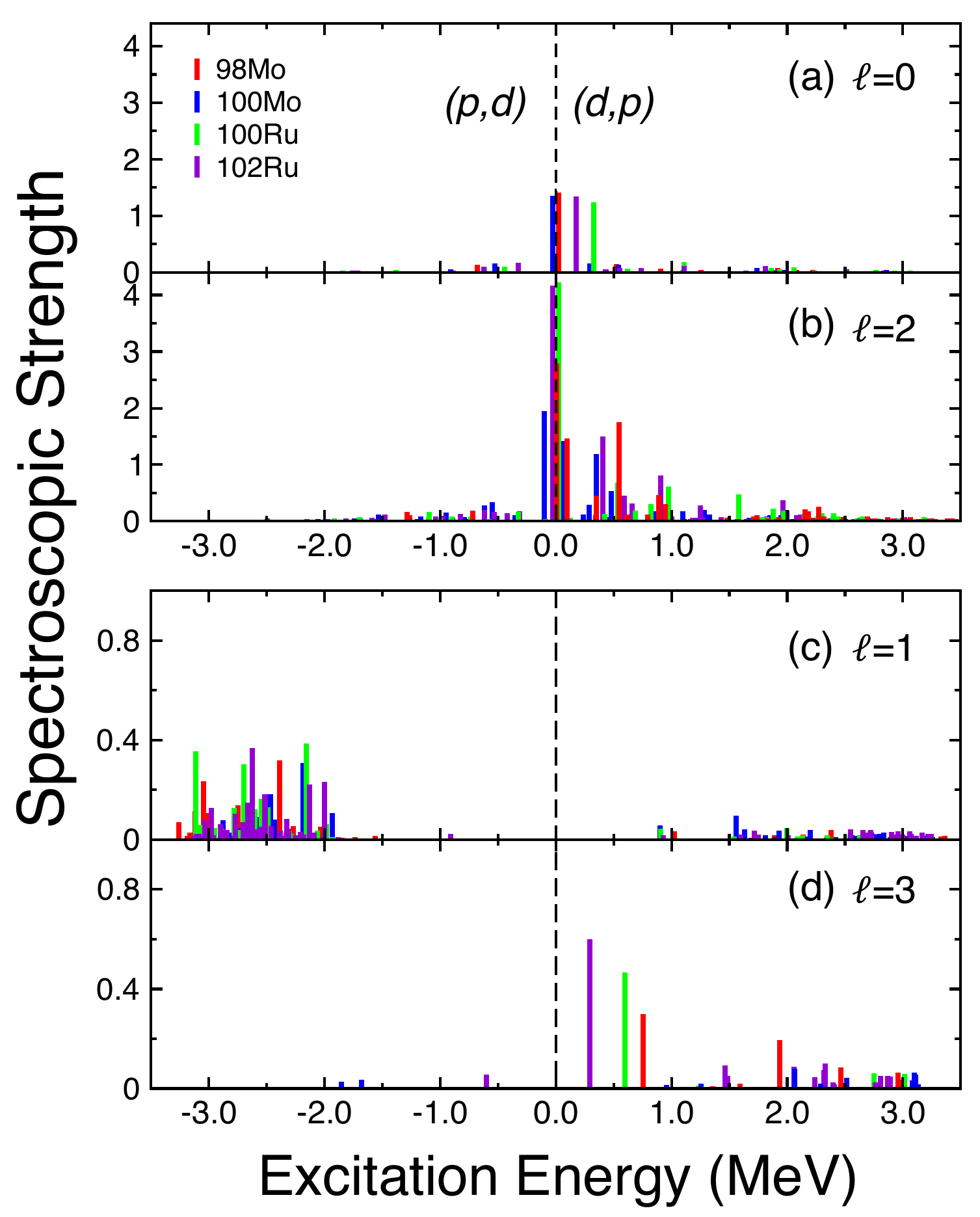}
\caption{\label{pddp} (Color online) Distributions of the spectroscopic strength of states populated in $(p,d)$ and $(d,p)$ neutron transfer reactions on targets of $^{102}$Ru (violet), $^{100}$Ru (green), $^{100}$Mo (blue) and $^{98}$Mo (red) as a function of excitation energy for  (a) $\ell=0$, (b) $\ell=2$, (c) $\ell=1$ and (d) $\ell=3$ transfers.  Note the difference in the vertical scales of the strength distributions for the valence orbitals in (a) and (b) compared to those of the out-of-shell strengths in (c) and (d). For the purposes of the figure,  strengths are plotted at negative energies for the ($p$,$d$) reaction and at positive energies for the ($d$,$p$) reaction. The strength of individual states has been obtained from the measured cross sections using the DWBA reaction modeling and normalization procedures described in Section III. For clarity, the strengths for the ground-state transitions in the two reactions have been combined and shifted slightly in excitation energy from zero.}
\end{figure*}

\begin{figure*}
\centering
\includegraphics[scale=0.6]{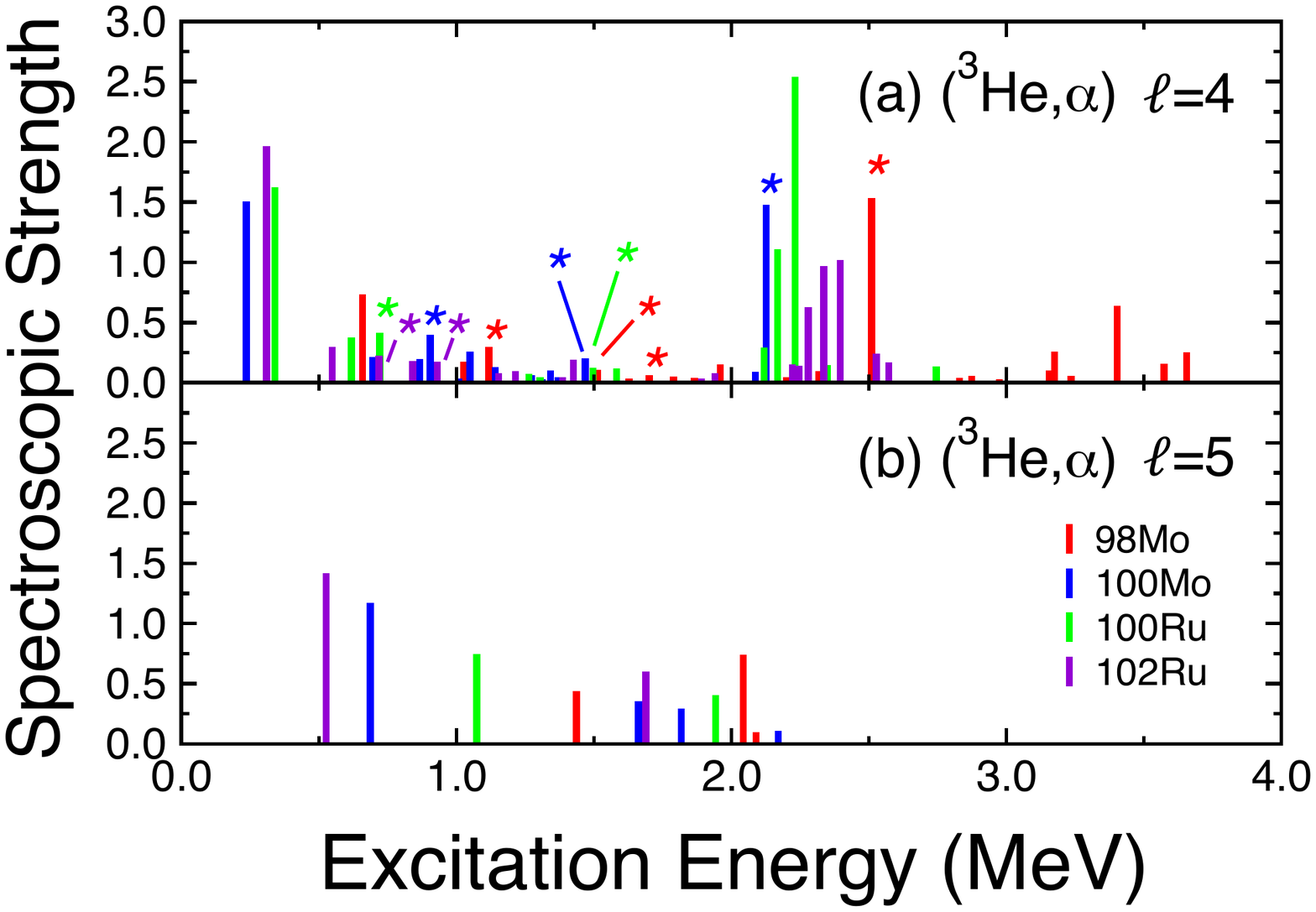}
\caption{\label{ha} (Color online) Distributions of the spectroscopic strength of states populated in  ($^3$He,$\alpha$)  neutron transfer reaction on targets of $^{102}$Ru (violet), $^{100}$Ru (green), $^{100}$Mo (blue) and $^{98}$Mo (red) as a function of excitation energy for  (a) $\ell=4$ and (b) $\ell=5$ transfers. The strength of individual states has been obtained from the measured cross sections using the DWBA reaction modeling and normalization procedures described in Section III. The asterisks indicate states with a $J^\pi=9/2^+$ assignment in the literature. Some states have been displaced slightly from their true excitation energy for clarity.}
\end{figure*}

It is instructive at this point to consider  the distribution of transfer strength in the residual nuclei. Figs.~\ref{pddp} and \ref{ha} show the distributions of spectroscopic strength defined as the spectroscopic factor $C^2S$ for removal reactions or $(2j+1)C^2S$ for addition reactions. (The spectroscopic factors have been obtained using the DWBA modeling and reaction normalization discussed in detail in Section III and are available as part of the Supplementary Material~\cite{Supplemental}.) 

Fig.~\ref{pddp} shows the distribution of spectroscopic strength for low $\ell$ transfers obtained from the ($d$,$p$) and ($p$,$d$) reactions as a function of excitation energy, where strength associated with  states populated in the latter reaction is plotted at negative excitation energies. Considering first the strength distributions for valence orbitals, the $\ell=0$ distributions are approximately Lorentzian in form with a centroid close to zero and a width of around 100~keV.  The $\ell=2$ strength is similarly centred at low excitation energies. Not all the states with $\ell=2$ have a firm $J^\pi$ assignment in the literature, but many of the stronger states at low excitation do have information on the spin quantum numbers. For example, the strong states clustered around 0 MeV in Fig.~\ref{pddp} are $5/2^+$ states. Most of the states with a strength greater than 0.5 at energies above  250~keV have $3/2^+$ assignments, where $J^\pi$ assignments are available. This is qualitatively consistent with the  energetic ordering of the $d_{5/2}$ and $d_{3/2}$ orbitals. Rough estimates of the unobserved strength were obtained in the following way. Lorentzian curves where fitted to the data and  the area under these fits outside of the excitation energy range of the measurements was only $\sim$2 to 3\% of the total, suggesting that the majority of the low-lying strength of the $s_{1/2}$ and $d$ orbitals has been captured in the data. Such estimates are consistent with similar studies that  have been performed \cite{Ni}. 

The out-of-shell strength distributions are somewhat different in character and weaker in overall strength; note the difference in the scale of the vertical axes for Fig.~\ref{pddp} (a) and (b) compared with Fig. ~\ref{pddp} (c) and (d). The $\ell=1$ strength, shown in Fig.~\ref{pddp}(c), appears at higher energies in both reactions, consistent with the tails of strength distributions from the next oscillator shells above and below the valence orbitals.  

The $\ell=3$ strength (see Fig.~\ref{pddp}(d)) is similarly weak and mostly at high excitation in the $(d,p)$ reaction, constituting a tail of strength from the shell above. There are single low-lying states populated by the $(d,p)$ reaction in $^{99}$Mo,  $^{101}$Ru and $^{103}$Ru with spectroscopic strengths up to $\sim$0.6; these states have also been observed in previous work, for example \cite{101Ru_dp1, 103Ru_dp3, Ru_dt}. Low-lying $\ell=3$ strength of this magnitude, associated with the $1f_{7/2}$ orbital from the shell above, has been predicted by modeling these transitional systems as a single neutron outside a weakly prolate core (see detailed discussion in Ref.~\cite{101Ru_dp1} and references therein).  In the $(p,d)$ reaction, $\ell=3$ strength is limited to a small number of very weakly populated states lying below 2~MeV. No strength has been identified with $^{100}$Ru and $^{98}$Mo targets,  a single state with spectroscopic strength of 0.05 in  $^{101}$Ru and two rather tentative  $\ell=3$ transitions in $^{99}$Mo, each with strength less than 0.01, have been found. These observations put a limit on the occupancy of $1f$ orbitals in the ground states of the target nuclei. It appears that the occupancy of the $1f$ orbital in these nuclei is $\lesssim 0.05$ neutrons, while the $0f$ shell is well below the Fermi surface.

Fig.~\ref{ha} shows a similar plot of spectroscopic strength for higher $\ell$ transfers taken from the ($^3$He,$\alpha$) reaction. The $\ell=5$ strength is confined to a small number of states at excitation energies in each residual nucleus at or below $\sim$2~MeV. The $\ell=4$ strength distribution is somewhat different with a number of strong states at low  energy, then the strength falls with increasing excitation  until some more prominent $\ell=4$ peaks are encountered above 2~MeV. This is consistent with an overall picture of low-lying $\ell=4$ strength associated with the valence $g_{7/2}$ orbital, but the presence of the deeper lying $g_{9/2}$ state at higher excitation. Indeed, below 2~MeV, all the states with spectroscopic strengths larger than 0.4 have been assigned as $J^\pi ={7/2}^+$ in the literature \cite{A=97,A=99,A=101}, although some weak $9/2^+$ states are also present in the same energy region. States whose spins are known to be $9/2^+$ are indicated by an asterisk in Fig.~\ref{ha}, although above 2~MeV the spins of most of the states are unknown. However, in $^{97}$Mo, the strong state at 2.510~MeV was assigned as $J^\pi=9/2^+$ from analysing powers measured in a $(d,t)$ reaction \cite{Mo_Pol_dt}. The lack of a complete set of $J^\pi$ assignments introduces some problems for the current work in disentangling $g_{7/2}$ and $g_{9/2}$ strengths. A choice was made to associate all $\ell=4$ strength below 2~MeV that does not have a previous $J^\pi=9/2^+$ assignment  with the $g_{7/2}$ orbital. Clearly other choices might be made in the absence of new spin assignments, which introduces a systematic error in the final occupancy analysis that will be discussed below.   To place this choice on a more quantitative footing, more than 90\% of the strength associated here with the $g_{7/2}$ orbital is in states with existing ${7/2}^+$ assignments.

\begin{figure}
\centering
\includegraphics[scale=0.4]{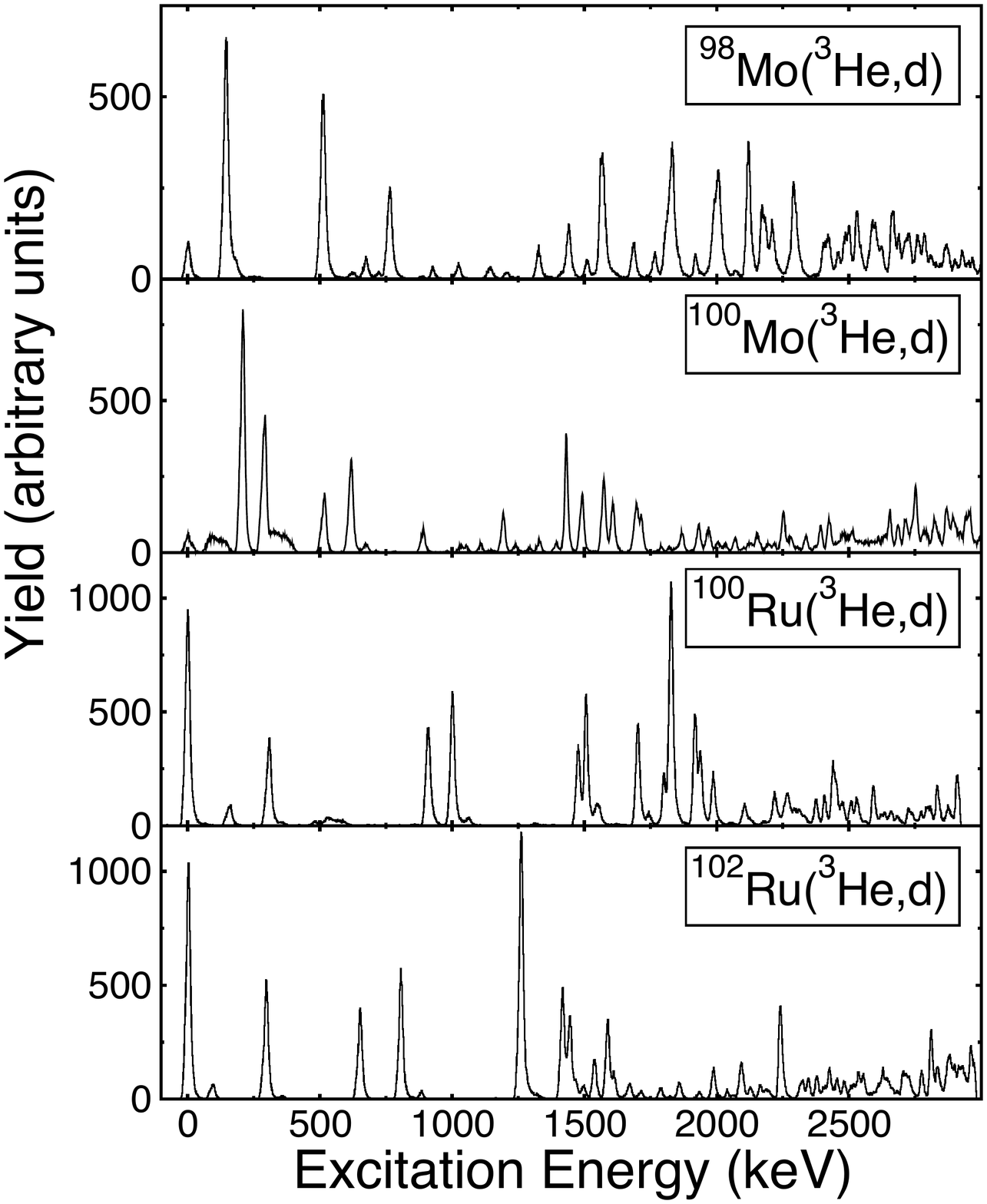}
\caption{\label{hd} Spectra of deuterons from the ($^3$He,$d$) reaction on targets of  $^{98}$Mo, $^{100}$Mo, $^{100}$Ru and $^{102}$Ru  at a laboratory angle of 6$^\circ$ as a function of the excitation energy in the residual nucleus. }
\end{figure}

\begin{figure}
\centering
\includegraphics[scale=0.5]{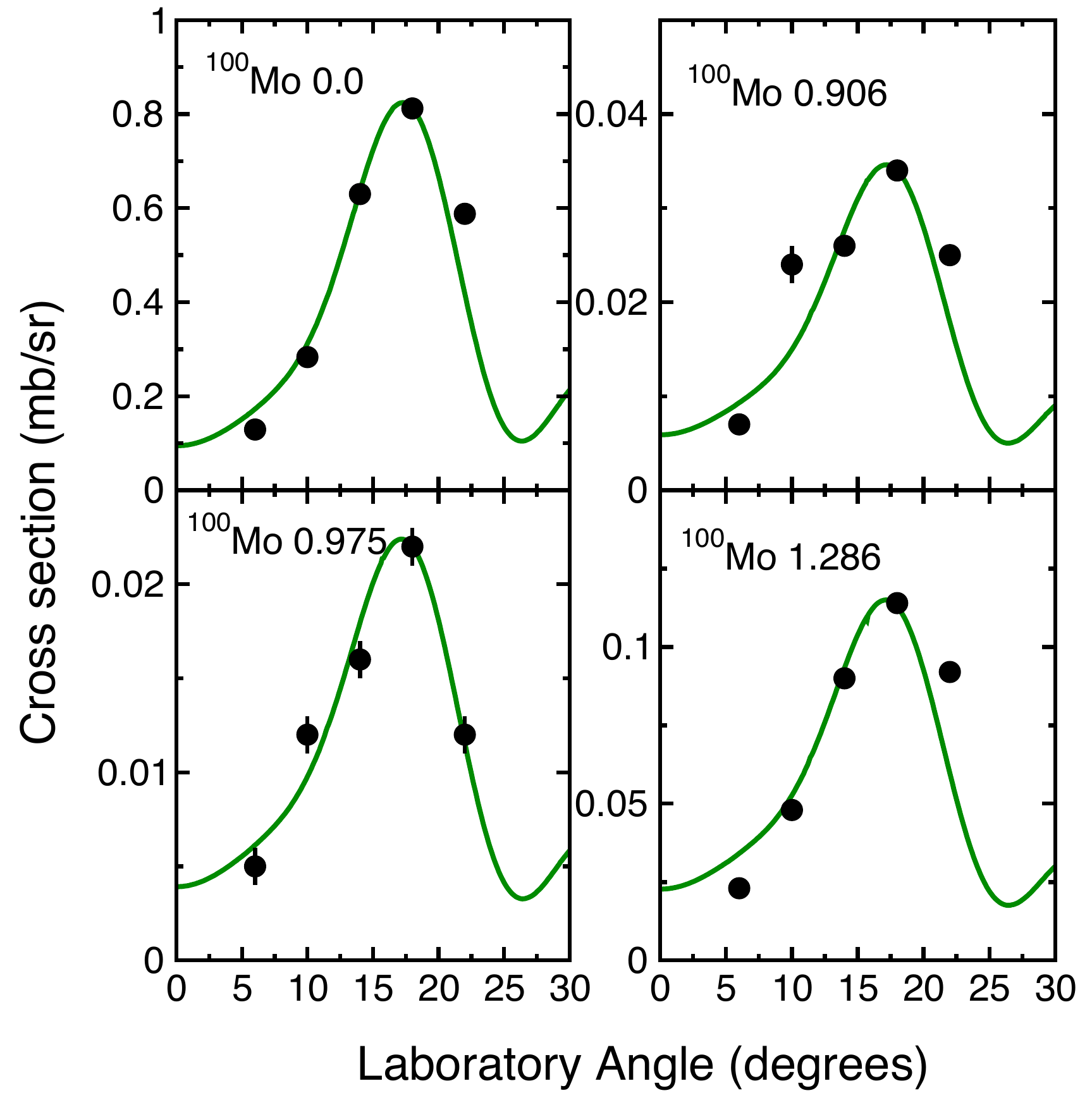}
\caption{\label{L=4}  (Color online) Examples of angular distributions for $\ell=4$ transitions assigned in the current work from the  $^{100}$Mo($^3$He,$d$) reaction and for the previously assigned  $\ell=4$  transition populating the residual ground state.  The data are compared to the results of DWBA calculations using parameters listed in Section III for $\ell=4$. The angular distributions are labelled by the target nucleus and the excitation energy in the residual system in units of MeV.}
\end{figure}

\begin{figure*}
\centering
\includegraphics[scale=0.6]{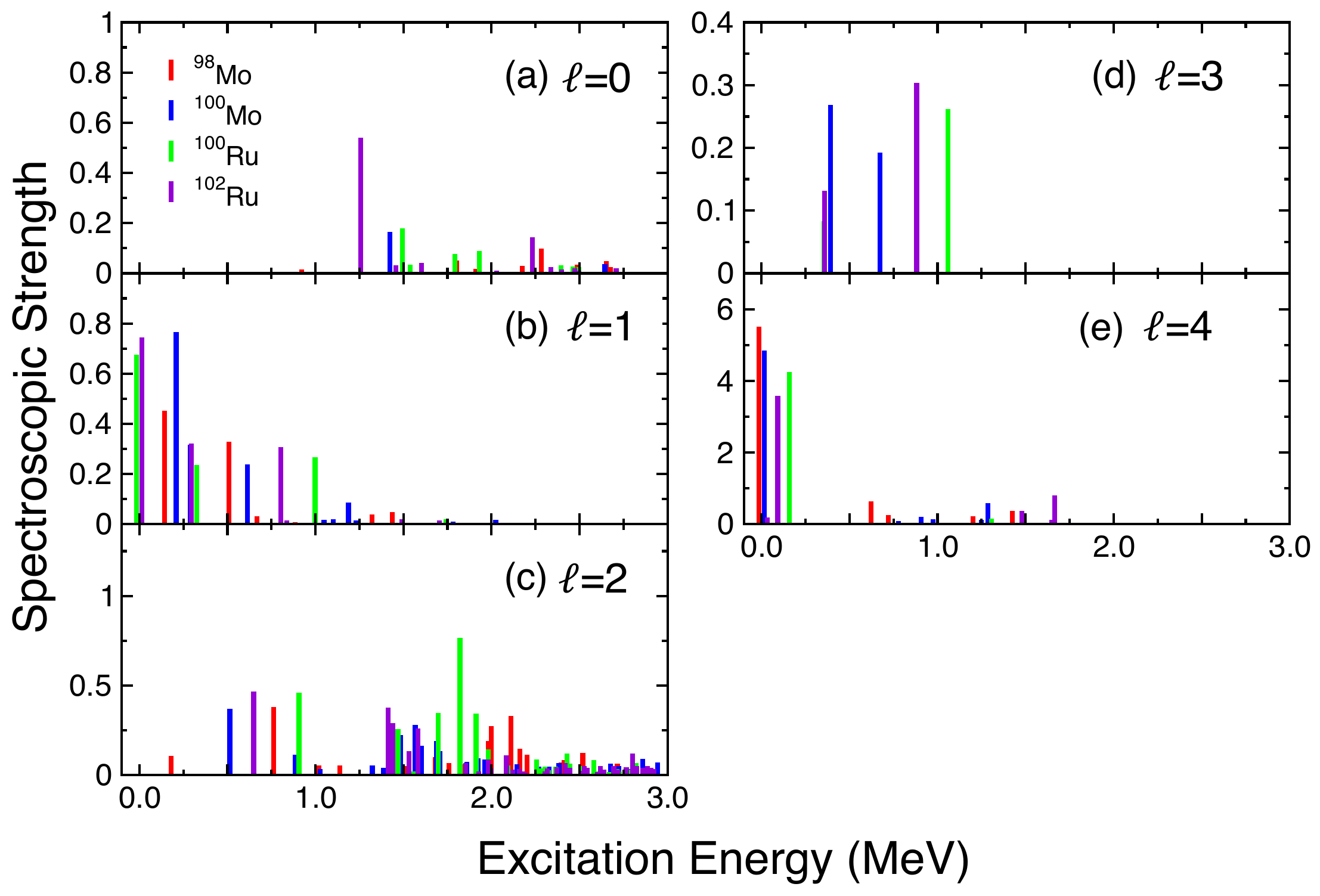}
\caption{\label{proton_strength} (Color online) Distributions of the relative strength of proton states populated in the ($^3$He,$d$)  reactions on targets of $^{102}$Ru (orange), $^{100}$Ru (green), $^{100}$Mo (red) and $^{98}$Mo (blue) as a function of excitation energy for (a) $\ell=0$, (b) $\ell=1$, (c) $\ell=2$, (d) $\ell=3$ and (e) $\ell=4$ transitions.  The relative strength of individual states  has been obtained from the measured cross sections using the DWBA reaction modeling and normalization procedures described in Section III. The strength associated with the population of the ground states have been displaced in some cases slightly from 0 MeV for clarity.}
\end{figure*}

\subsection{Proton Transfer Reactions}

The ($^3$He,$d$) proton-adding reactions were initiated using beams  at an energy of 36 MeV.   Data were recorded up to excitation energies of at least 2.7~MeV,  performed at one magnet setting. Excitation energies of states in the residual nucleus were obtained by comparison with previously observed states taken from Refs.~\cite{A=97,A=99,A=101} and some representative spectra are shown in Fig.~\ref{hd}. The excitation energies obtained were generally measured to better than 3~keV, although in the ruthenium targets this rises to 10~keV at the highest excitations measured as there are fewer previously known states for calibration.  Typical energy resolutions of 20~keV FWHM were obtained and measurements were made at a series of angles listed in Table~\ref{Ang-Table}.

The assignments of $\ell$ transfer were checked using angular distributions and Fig.~\ref{Dists} shows some examples. There are no previously reported data for this reaction on ruthenium targets in the literature, although nearly all of the states carrying strength from the valence nucleon orbitals have  assignments deduced by other types of measurement \cite{A=97,A=99,A=101}. In total, 92\% of the strength used in deducing the proton occupancies is associated with the population of states with  previous assignments; across the individual targets used, the  percentage of strength with previous assignments are 99\%, 82\%, 97\% and 93\% for $^{98}$Mo, $^{100}$Mo, $^{100}$Ru and $^{102}$Ru, respectively. In the reactions on $^{100}$Mo, the new assignments made here were predominately  $\ell=4$ states. Some examples of the relevant angular distributions are compared to that for the known $\ell=4$ ground-state transition and to DWBA predictions in Fig.~\ref{L=4}.

The distributions of spectroscopic strength $(2j+1)C^2S$ for proton addition  obtained using the ($^3$He,$d$) reaction are shown in Fig.~\ref{proton_strength}(a)--(e). The  transfer associated with the proton valence orbitals has $\ell=1,$ 3 and 4. With increasing excitation energy, the $\ell=1$ strength  falls off rapidly  and is contained mostly in the first 1.5~MeV as shown in Fig.~\ref{proton_strength}(b). There is not much $\ell=3$ strength, all of which lies at energies below 1.1~MeV; no $\ell=3$ transitions were apparent in reactions on the $^{98}$Mo target. For $\ell=4$, the majority of the strength identified in the ($^3$He,$d$) reaction is in a single low-lying 9/2$^+$ state below 0.5~MeV in each residual nucleus, with some weaker fragments at energies up to 1.5~MeV (see Fig.~\ref{proton_strength}(e)).

The  distributions of strength associated with non-valence orbitals with $\ell=0$ and 2 (see Fig.~\ref{proton_strength}(a) and (c)) cover higher excitation energy regions compared to the valence strengths. For example, the distribution of $\ell=0$ strength (see Fig.~\ref{proton_strength}(a)) appears above  1~MeV, consistent with a tail of relatively weak strength from the shell above the valence orbitals. Similarly, much of the $\ell=2$ strength lies in many small fragments at higher excitations. There is some $\ell=2$ strength that appears in a number of individual states at energies less than 1~MeV that have been interpreted previously by core-coupling \cite{Bindal1} and Coriolis-coupling \cite{Bredbacker} models, where $2d$ strength is brought down in excitation energy, with spectroscopic strengths similar to those observed here, by a mechanism somewhat analogous to the low-lying $\ell=3$ neutron strength discussed above. 

Proton-removal reactions were not studied in the current work due to limitations in the available beam energy for ($d$,$^3$He) reactions and difficulties with tritium handling for ($t,\alpha$) reactions. However limited information is available in the literature, albeit only on molybdenum isotopes, which can be used to assess the contributions to non-valence-shell orbitals in the ground states. A study of the $(d,^3$He) reaction \cite{Bindal2} has been performed and  polarized ($t$,$\alpha$) data is reported in Ref.~\cite{Flynn}. Neither reaction on $^{98,100}$Mo targets populated any $\ell=0$ strength. There are some inconsistencies between these two studies concerning $\ell=2$ strength, which are likely attributable to the lower resolution of the $(d,^3$He) measurement. In $^{97}$Nb, the $(d,^3$He) work observed states with $\ell=2$ strength at 1.764 and 2.090~MeV extracted by fitting several states to broad multiplet peaks; the ($t$,$\alpha$) study had higher resolution, made different assignments and reported no $\ell=2$ population in this nucleus.  A state was observed at 0.817~MeV in both $(d,^3$He) and ($t$,$\alpha$) reactions, the latter also populated  states in $^{99}$Nb at 0.469 and 0.763 MeV, all with {\it tentative} $\ell=2$ assignments.  Using the DWBA prescription presented below and cross section data from these references, the  spectroscopic factors for the 0.469-, 0.763- and 0.817-keV states were estimated to be 0.09, 0.04 and 0.11. This allows us to estimate a limit for the occupation of $\ell=2$ in the ground state of $^{99}$Nb at the level of at most $\sim$0.1 protons.

\section{DWBA Modeling and Normalization}

Spectroscopic factors were deduced from the experimentally measured cross sections by comparison with the results of calculations using the distorted-wave Born approximation performed with the finite-range code {\sc ptolemy} \cite{ptolemy}. The optical potentials and bound states used in these calculations were chosen to be consistent with  a recent global analysis of the quenching of spectroscopic strength \cite{quenching} and are summarized below.

The form factors associated with the light-ion wave functions were taken from recent microscopic calculations. Those for the deuteron in ($d$,$p$) and ($p$,$d$) reactions were deduced using the Argonne $v_{18}$ potential \cite{wiringa}. Recent Green's function Monte Carlo calculations provided form factors for $A=3$ and $A=4$ species \cite{Brida}.

The single-particle wave functions of the transferred particle in the heavy bound state were generated using a Woods-Saxon potential with fixed geometric parameters: radius parameter $r_0=1.28$~fm and diffuseness $a=0.65$~fm. The depth was chosen to reproduce the measured binding energies. A spin-orbit component based on the derivative of a Woods-Saxon form with a geometry defined by $r_{\rm so}=1.10$~fm and $a_{\rm so}=0.65$~fm, with a depth $V_{\rm so}$ of 6~MeV was used.

The distortions of incoming and outgoing partial waves were described using global optical-model potentials for protons, deuterons, helions and tritons taken from Refs.~\cite{protons, deuterons, helions}. An $\alpha$ potential deduced from elastic scattering in the $A=90$ region \cite{basani} was used.

In order to best satisfy the approximations of the DWBA approach, spectroscopic factors were deduced from cross sections at angles closest to the first peak of the angular distributions.  In neutron transfer, the ($d$,$p$) and ($p$,$d$) reactions were used to determine spectroscopic strength for the lower orbital angular momentum transfer, $\ell=0$ and $2$, and that for $\ell=4$ and $5$ were deduced from ($^3$He,$\alpha$) in order to ensure optimal momentum matching. The ($^3$He,$d$) reaction is reasonably well-matched for all the relevant $\ell$ in proton transfer.

The DWBA calculations used to extract spectroscopic factors from experimental cross sections carry an uncertainty in overall absolute normalization. Methods for determining the value of this normalization have been developed using the  Macfarlane-French sum rules  \cite{macf} that associate the summed spectroscopic strength to occupancies and vacancies of nucleon orbitals. Consistent results can be obtained by adopting a systematic approach to this process (see for example Ref.~\cite{Ni}). If  the total low-lying strength is normalized to the full independent-particle value, the degree to which the  resulting normalization factor deviates from unity is related to the quenching of  single-particle strength that has been observed in other types of reactions such as $(e,e^{\prime}p)$. Here we follow methods of Ref.~\cite{quenching} where a large-scale analysis resulted in normalization factors that were quantitatively consistent with previous measurements of such quenching. 
 
\begin{table}[H]
\caption{\label{N-Table} Normalization factors for the DWBA calculations obtained using  procedures described in the text.\\}
\begin{ruledtabular}
\begin{tabular}{r c c c r}
  &($d,p$)/($p,d$)&($d,p$)/($p,d$)&($^3$He,$\alpha$)&($^3$He,$d$)\\
 & $\ell=0$&$\ell=2$ &$\ell=4$ and 5 &\\
\hline
\\
$^{102}$Ru	& 0.642& 	0.673&	0.570&	0.682\\
$^{100}$Ru	& 0.610&		0.555&	0.572&	0.647\\
$^{100}$Mo	& 0.624&		0.617&	0.576&	0.639\\
$^{98}$Mo	& 0.595&		0.612&	0.538&	0.622\\\\
\hline
\\
Mean 		& 0.618	&	0.614&	0.564&	0.647\\
St Dev & 0.020 & 0.048&0.018&0.025\\
\\
 \end{tabular}
 \end{ruledtabular}
 \end{table}

For neutron-transfer reactions the following normalization procedure was adopted. The first step was to use the ($d$,$p$) and ($p$,$d$) data  to deduce the summed spectroscopic strength for $\ell=0$ and 2,  associated with the $2s_{1/2}$ and $1d$ orbitals. Via the sum rules \cite{macf}, the summed strength for the neutron-adding reaction is proportional to the vacancy in the associated orbital. Similarly for the neutron-removing reaction, the summed strength  is proportional to the occupancy. A DWBA normalization was chosen such that the overall sum of strength from both neutron addition and removal gives the orbital degeneracy. Initially, this was done separately for both $\ell=0$ and 2 and  for reactions on each target. The resulting normalization factors are shown in Table~\ref{N-Table}. The average normalization across all targets  for $2s_{1/2}$ transfer was found to be 0.618 and that for the combined strengths associated with $1d$ orbitals was 0.614. The individual normalization values varied across the targets used by  3\% and 8\% for $\ell=0$ and $\ell=2$ respectively. This variation,  and that between the two $\ell$ transfers, is small and so the overall average normalization constant of 0.616 was used in the subsequent analysis.  

Assuming that the $N=50$ shell is closed, valence neutrons only occupy the $2s_{1/2}$, $1d$, $0g_{7/2}$ and $0h_{11/2}$ orbits (see comments above about the validity of this assumption). A normalization for the ($^3$He,$\alpha$) reaction was deduced by requiring that the sum of the previously normalised spectroscopic strength  from ($p$,$d$) data for $\ell=0$ and 2 (i.e. the occupancy of those orbitals) and the spectroscopic strength for $\ell=4$ and 5 states from the ($^3$He,$\alpha$) reaction results in the expected total number of valence nucleons. The average normalization for $\ell=4$ and 5 transitions in the ($^3$He,$\alpha$) reaction was found to be 0.564, with a 3\% variation across the four targets. 

For proton transfer, a similar procedure was used where the total spectroscopic strength populated using the ($^3$He,$d$) reaction for all states corresponding to valence protons was required to equal the expected number of proton vacancies in the $Z=50$ shell. The resulting normalization factor was 0.647 and the variation across targets was 4\%.

A substantial set of transfer data was analyzed recently in a consistent fashion to determine the normalization of DWBA calculations and the associated quenching factor for single-particle motion in near-stable nuclei \cite{quenching}. That analysis indicated that spectroscopic factors for a variety of light-ion induced transfer reactions across targets from $^{16}$O to $^{208}$Pb are quenched with respect to values from mean-field theory by a factor of 0.55, with a root-mean square spread of 0.1. This compares favorably with the normalization factors deduced in the current work. The consistency with independent data sets, along with the consistency across all four targets and between the different $\ell$ values,  gives confidence in the  methodology used. 

When considering isospin effects in the reactions, it should be noted that neutron adding and proton removal result in the population of states with a single value of isospin $T+1/2$, where $T$ is the isospin of the target. In contrast, in proton adding and neutron removal, states with both $T+1/2$ and $T-1/2$ are accessible. The set of states with  higher isospin lie at higher excitation energies and are not observed in the kind of experiment described here. However, the summation in the Macfarlane and French sum rules should, in principle, contain strength associated with both values of isospin  and the normalization procedure described above needs correcting for the unobserved strength. Using isospin symmetry, this could be done for proton-adding/neutron-removal reactions using spectroscopic strengths associated with the same orbitals populated in neutron-adding/proton-removal reactions \cite{Schiffer}. However, protons and neutrons in these nuclei reside in different oscillator shells and the valence orbitals populated in neutron removal from all the target nuclei considered here are empty of protons. Subsequently, the required proton-removal strength is small. Given that the  isospin Clebsch-Gordan coefficient is also small, the correction for the unobserved higher isospin is smaller still. Similarly for proton addition, the expectation is that the neutron-adding spectroscopic factors for $g_{9/2}fp$ orbits would be small due to their high occupancy. Indeed, even if all the observed ($d$,$p$) strength for $\ell=1$, $3$ or $4$ observed here were associated with the $1p$, $0f_{5/2}$ and $0g_{9/2}$ orbitals, which is clearly a gross over-estimate, the normalization factors only change by a few percent. The isospin corrections were therefore considered small, compared to other uncertainties, in the current work and were not applied to the final analysis.

\section{Nucleon Occupancies}

\begin{figure*}
\centering
\includegraphics[scale=0.5]{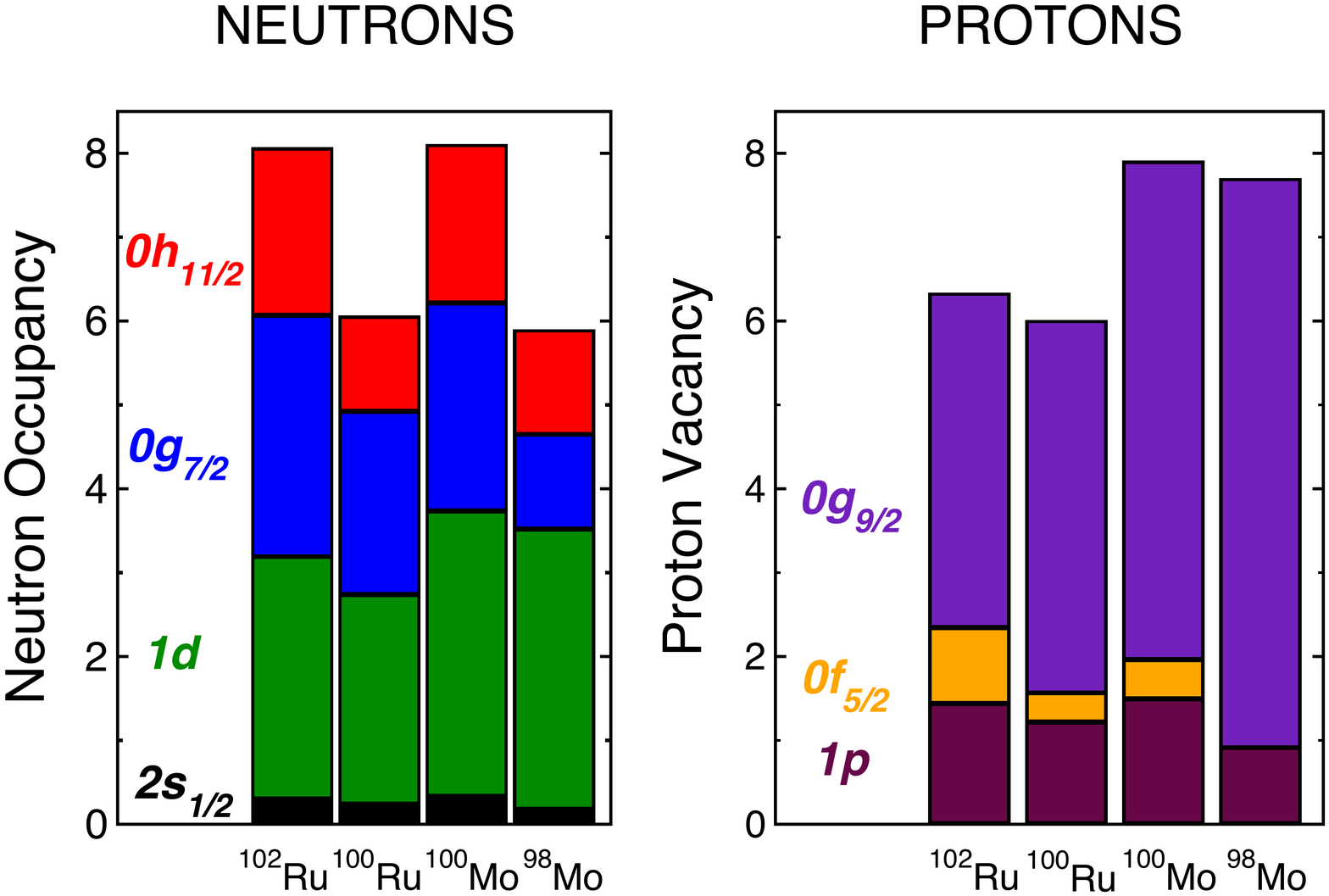}
\caption{\label{O} (Color online) Experimentally determined neutron occupancies and proton vacancies for the valence orbits in $^{100}$Mo and $^{100}$Ru, along with $^{102}$Ru and $^{98}$Mo which are used for consistency checks. }
\end{figure*}

Nucleon occupancies were deduced from  summed spectroscopic strengths determined using the normalization factors described in the previous section. The neutron occupancies were extracted from the neutron-removing reactions and are listed in Table~\ref{ON-Table}. Proton vacancies obtained from the ($^3$He,$d$) reaction are given in Table~\ref{OP-Table}.  These data are also shown graphically in Fig. \ref{O}.  As noted above, the occupancy of non-valence orbitals in the ground states of these nuclei is estimated to be lower than 0.1 nucleons.

\begin{table*}
\caption{\label{ON-Table} Experimental neutron occupancies determined from neutron-removing reactions. The difference between the summed occupancy and the expected number of valence neutrons is also given. The decreases in neutron occupancies of each orbital associated with  double $\beta$ decay of $^{100}$Mo are  given at the bottom of the table. The errors quoted are based on relative variations due to choices of potentials in the DWBA and, in the case of high-$\ell$ orbitals, a contribution to reflect a systematic effect from spin assignment for $\ell=4$ (see text for details). \\}
\begin{ruledtabular}
\begin{tabular}{r c c c c c c r}
  &$2s_{1/2}$&$1d$&$0g_{7/2}$&$0h_{11/2}$&Total&Expected&Difference\\
\hline
\\
$^{102}$Ru	&0.29(1)	&2.89(14)		&2.88(38)	&2.00(14)		&8.05(43)	&8	&0.05\\
$^{100}$Ru	&0.23(1)	&2.50(12)	 	&2.19(15)	&1.13(8)		&6.05(21)	&6	&0.05\\
$^{100}$Mo	&0.33(2)	&3.40(17)		&2.48(19)	&1.89(13)		&8.09(29)	&8	&0.09\\
$^{98}$Mo	&0.17(1)	&3.34(17)		&1.13(6)	&1.25(9)		&5.88(20)	&6	&-0.12\\
\\
$^{100}$Mo-Ru &0.09(2)	&0.90(21)	&0.30(24)	&0.76(15)	&2.05(36)& \\
 \end{tabular}
 \end{ruledtabular}
 \end{table*}
  

\begin{table*}
\caption{\label{OP-Table} Experimental proton vacancies determined from the ($^3$He,$d$) reaction. The difference between the summed vacancy and the expected number of valence proton holes is also given. The increases in proton occupancy in each orbital associated with  double $\beta$ decay of $^{100}$Mo are also given at the bottom of the table. The errors quoted are based on relative variations due to choices of potentials in the DWBA (see text for details).\\ }

\begin{ruledtabular}
\begin{tabular}{r c c c  c  c r}
  &$1p$&$0f_{5/2}$&$0g_{9/2}$&Total& Expected&Difference\\
\hline
\\
$^{102}$Ru	&1.43(7)	&0.90(5)	&3.98(20)		&6.32(22)	&6	&0.32	\\
$^{100}$Ru	&1.21(6)	&0.35(2)	&4.44(22)		&6.00(23)	&6	&0.00\\
$^{100}$Mo	&1.49(7)	&0.47(2)	&5.94(30)		&7.89(31) &8	&-0.11\\
$^{98}$Mo	&0.91(5)	&  -- 		&6.78(34)		&7.69(34) &8	&-0.31\\
\\
$^{100}$Mo-Ru &0.28(10)	&0.12(3)	&1.50(37)	&1.90(38)\\
 \end{tabular}
 \end{ruledtabular}
 \end{table*}

There are a number of systematic effects that could potentially influence the methodology adopted in deducing the nucleon occupancies. For example, it is well known that the results of DWBA calculations carry significant sensitivity to the input parameters used. The sensitivity of the current calculations was investigated using a variety of different optical-model potentials. Whilst the absolute values of the calculated cross sections varied considerably (by up to $\sim$20\%), the relative numbers relevant for the current analysis varied by up to 5\%. Since statistical contributions are generally small, this is the largest contribution to the uncertainty in the deduced orbital occupancies and has been used as a basis to estimate the errors quoted in Tables~\ref{ON-Table} and ~\ref{OP-Table}; the high-$\ell$ neutron occupancies have an additional contribution discussed below. Using these estimates, the combined error on the total number of valence particles inferred from the experiment is  typically $\sim$0.2$-$0.3 depending on  target. This is roughly consistent with the root-mean-square deviation  of this number from the expected number of valence particles across the targets, 0.1 for neutrons and 0.2 for proton holes. These error estimates are also similar to those obtained in occupancy measurements of other nuclear systems \cite{Ge_n, Ge_p, Te, Entwisle, Szwec}.

Beyond direct nucleon transfer, there are other more complicated reaction mechanisms that can contribute to the measured yields. Recent transfer work on nickel isotopes \cite{Ni} presented a method to estimate the contribution of multistep processes by comparing the spectroscopic strength of states populated by a well-matched and a poorly-matched reaction. This was applied to $\ell=4$ transitions in the current data set populated by the ($p$,$d$) and ($^3$He,$\alpha$) reactions and gave a very similar estimate to that in Ref.~\cite{Ni}. Multistep processes are estimated to contribute at a level of around $0.002(2j+1)$ in the spectroscopic strength of states deduced using a reaction with good matching. Most of the strength contributing to the sum-rule analysis is from states populated much more strongly than this level and therefore multistep processes appear not to influence the data strongly.

There are a few influences associated with spin assignments that could affect the deduced occupancies. The most important of these is the assignment of the spins of states populated via $\ell=4$ transfer in neutron-removal reactions. As noted above, a choice was made to associate all $\ell=4$ strength below 2~MeV with the $0g_{7/2}$ orbital unless it had a previous ${9/2}^+$ assignment, but other approaches could be adopted. For example, one could use only the strength associated with states with a previous ${7/2}^+$ assignment. If this were done, the $0g_{7/2}$ occupancies in the $A=100$ isotopes change by $\sim$0.1 neutrons due to changes in the summed $\ell=4$ strength in those nuclei, with a smaller 5\% decrease in $\ell=5$ occupancies due to the associated shift in the ($^3$He,$\alpha$) normalization. However, the consistency in the individual normalization factors is then worse than in the adopted approach, probably reflecting variation in the extent of $J^\pi$ assignments for the residual nucleus in the literature. These effects have been added in quadrature to the errors for $\ell=4$ and 5 orbitals in Table~\ref{ON-Table} as an estimate of this systematic effect. Variation in the excitation-energy limit used to exclude the higher-lying $0g_{9/2}$ strength has less consequence. 

In addition, there are a number of states observed in the ($^3$He,$\alpha$) reaction that are not obviously populated in the ($p$,$d$) reaction; these are candidates for $\ell=4$ or 5 transitions, but the lack of  ($p$,$d$) data makes assignment difficult and they have not been included in the analysis. If they were introduced, the {\it maximum} effect they make for the occupancies of the high-$\ell$ neutron orbitals is 0.1 nucleons. Other minor complications, such as tentative assignments and unresolved doublets, affect the final results at a much lower level.

\section{Discussion}

The measured neutron occupancies shown in Fig.~\ref{O} indicate that neutrons occupy each of the orbitals in the shell above $N=50$, with the different $\ell$ values full to at least 10\% of the maximum occupancy.  Although the current measurements  cannot distinguish between the two $1d$ orbitals, much of the $\ell=2$ strength populated here is associated with states that have a $J^\pi$ assignment in the literature (as summarized in Refs.~\cite{A=97,A=99,A=101,A=103} and references therein). The fraction of  $\ell=2$ strength without a $J^\pi$ assignment varies from  $\sim$10 to 25\% across the different targets. Using the known $J^\pi=5/2^+$ strength, a lower limit on the occupancy of the $1d_{5/2}$ orbital is estimated as 1.8, 1.9, 2.4 and 3.0 neutrons in $^{102,100}$Ru and $^{100,98}$Mo respectively, indicating that this orbital is responsible for most of the observed  $1d$ occupancy.

The proton Fermi surface lies below $Z=50$. The pattern of proton vacancy is shown in Fig.~\ref{O}, and illustrates that the $0f_{5/2}$ orbital is  almost full and the $1p$ orbitals carry around two thirds of their maximum occupancy, whereas the $0g_{9/2}$ state is only partially occupied. For the $\ell=1$ strength, at least 90\% of the states populated on each target have a $J^\pi$ assignment  in the literature (\cite{A=97,A=99,A=101,A=103} and references therein). Applying these $J^\pi$ assignments suggests that the $1p_{3/2}$ orbital has a vacancy of at most 14\% across the different targets, with  the $1p_{1/2}$ orbital empty to the level of at most 39\%. Given the vacancy in the $1p$ orbitals, it would appear from these results that the $Z=40$ sub-shell closure,  assumed in some shell-model calculations, is somewhat weak in these systems. 

\begin{figure}
\centering
\includegraphics[scale=0.6]{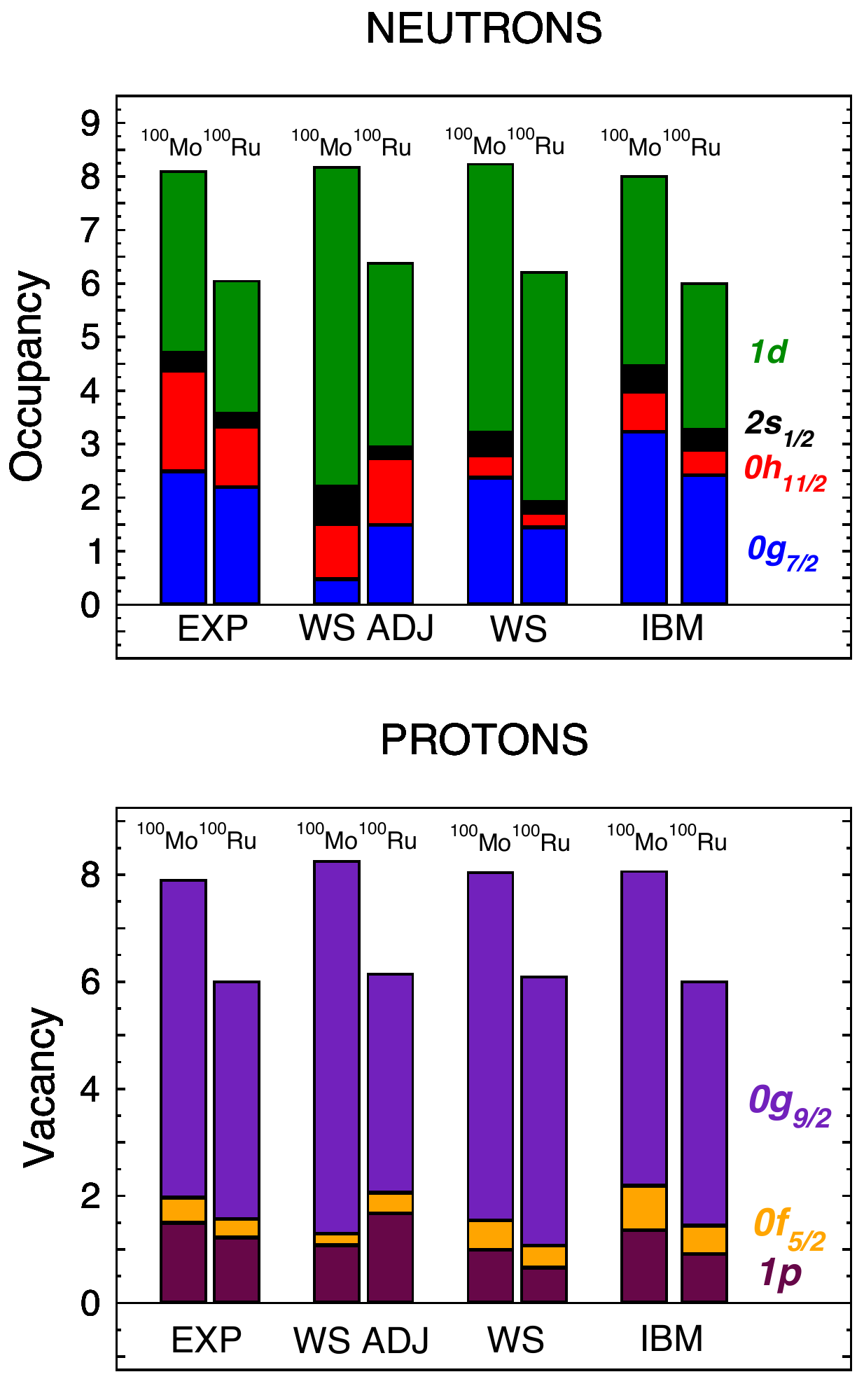}
\caption{\label{NO} (Color online) Experimentally determined neutron occupancy and proton vacancy for the valence orbits in $^{100}$Mo and $^{100}$Ru compared to those predicted by  the interacting boson model (IBM)~\cite{Kotila,Kotila16} and two different Woods-Saxon calculations ~\cite{Suhonen,Suhonen14}.}
\end{figure}

\begin{figure*}
\centering
\includegraphics[scale=0.9]{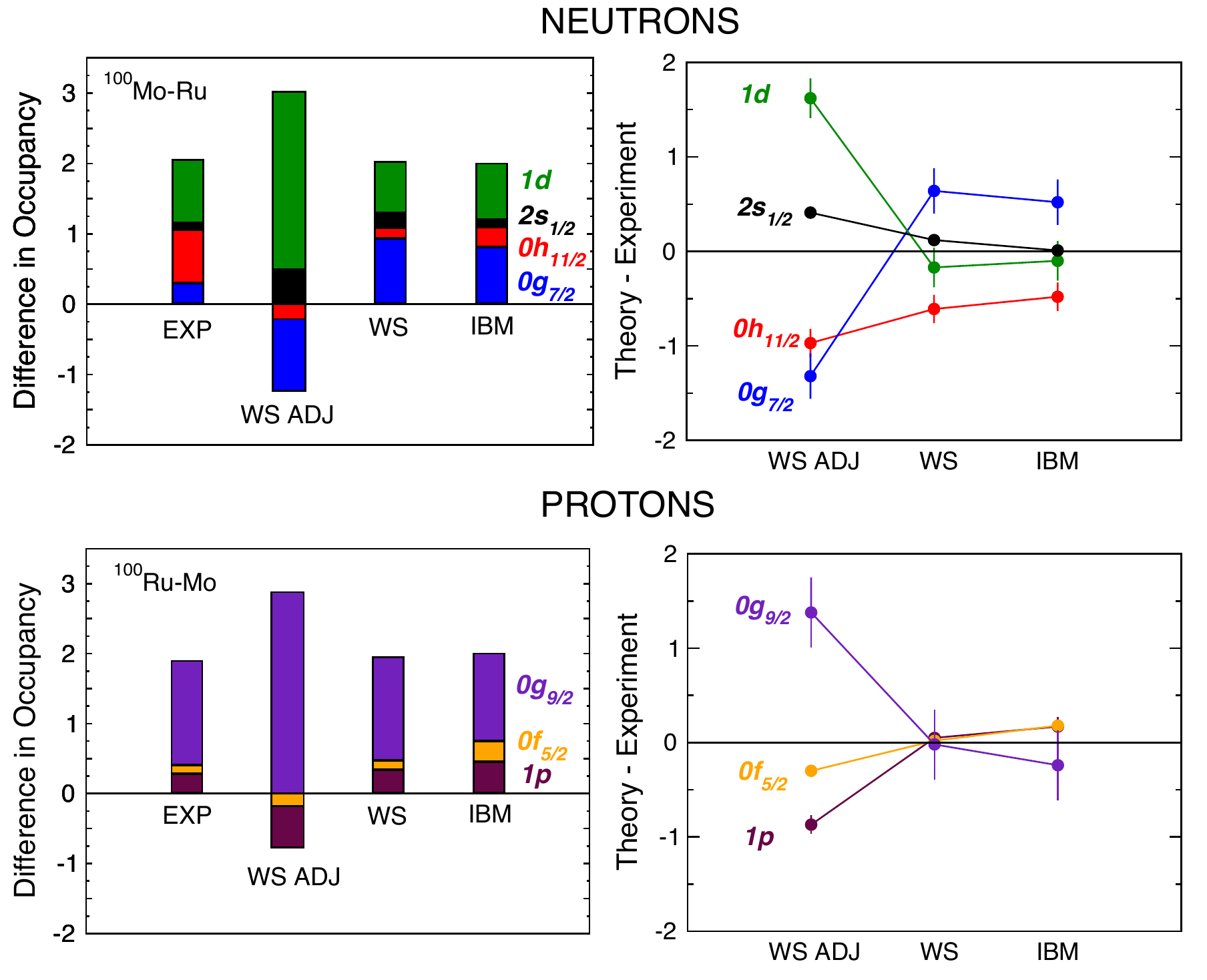}
\caption{\label{CO} (Color online) Left: Changes in the occupancy of valence nucleon orbitals during a double $\beta$ decay of $^{100}$Mo deduced from experimentally measured occupancies (EXP) compared to those predicted by a number of different theoretical calculations of  double $\beta$ decay where the the same labeling as Fig.~12 has been used (see text for details). The signs are chosen such that a reduction in the number of neutrons and a gain in the number of protons are positive numbers. Right: The difference between the theoretical calculations and experimental numbers plotted with experimental errors.   }
\end{figure*}

The comparison of measured nucleon occupancies with those extracted from theoretical studies of nuclear matrix elements for double $\beta$ decay has proved very instructive in the past, as illustrated by the example of Ref.~\cite{Simkovic} in the case of $^{76}$Ge decay. However, quantitative occupancy numbers are not always given in theoretical publications. 
The $^{100}$Mo$-^{100}$Ru system has been the subject of several theoretical determinations of the nuclear matrix element for $0\nu2\beta$ decay and associated orbital occupancies are available for calculations using the interacting boson model (IBM)  and quasi-particle random-phase approximation (QRPA). Nucleon occupancies can be extracted from the IBM wave functions relatively easily as discussed in Ref.~\cite{Kotila16}.  QRPA calculations take as input single-particle energies and occupancies, often from BCS calculations using a Woods-Saxon potential (WS). Whilst it is easy to use such inputs to compare with measured occupancies, it would be more consistent to compare the current results with the occupancies contained in the correlated QRPA ground states. This results in complications as standard QRPA methods do not automatically conserve particle number, even on average. Reformulations of QRPA methods that ensure average particle number conservation do exist; for example, the self-consistent renormalised approach (SRQRPA) taken in Ref.~\cite{Simkovic} has been applied to the $^{76}$Ge $0\nu2\beta$ decay system. There are differences in  occupancies predicted by the BCS approximation and SRQRPA, but these tend to be small except for some orbitals with higher orbital angular momentum \cite{Suhonen14}. Since, to the best of our knowledge, SRQRPA calculations have not been done for the $A=100$ system, here we will compare with the  available occupancies used as inputs to QRPA calculations and note this issue for future theoretical attention.

Valence neutron occupancies and proton vacancies are shown in Fig.~\ref{NO} compared to IBM and WS calculations. Two sets of results for  the Woods-Saxon potential are shown; one taken from a standard parameterization adopted near the line of stability \cite{Suhonen} (labeled WS in Fig.~\ref{NO}) and one (labeled WS ADJ in Fig.~\ref{NO}) after   adjustments to better reproduce quasi-particle states in nearby odd-A nuclei (see Ref.~\cite{Suhonen14} and references therein for details). This set has been used as input not only to calculations of both single EC, single $\beta$  and two-neutrino double $\beta$ decays \cite{Suhonen14,Suhonen02}, but has also been  used  for  $0\nu2\beta$ decay \cite{Suhonen02, Suhonen94}. 

For protons, most of these calculations appear to give a reasonable overall  description of the measured vacancies. For the IBM calculations, the discrepancies are at the level of a couple of tenths of a nucleon and probably within the uncertainties in the experiments. For the  WS results, the overall picture is similar, but discrepancies are slightly larger. However, in the case of the adjusted Woods-Saxon calculations, the comparison with the experimental vacancies is  worse than the other calculations, particularly for $^{100}$Mo where there is significant over prediction of the vacancy of the $0g_{9/2}$ orbital.

For neutrons, the comparisons are more mixed. The IBM calculations appear to slightly overestimate the neutron occupancy of the positive-parity orbitals at the expense of the  $0h_{11/2}$ orbit, which is predicted to have significantly lower occupation than the current data suggests. The underestimation of the occupancy of this intruder orbit persists in the WS calculations, but results in over prediction for $1d$ neutrons. The adjusted WS calculations do have a better reproduction of the experimental $0h_{11/2}$ occupancy, but fail to reproduce the numbers of neutrons in  the $1d$ and $0g_{9/2}$ orbitals; these discrepancies appear to be more dramatic in the case of $^{100}$Mo. The larger discrepancies referred to here are significant compared to the experimental uncertainties, accompanied by less significant issues with  $2s_{1/2}$ neutrons. None of the calculations fare as well with the neutron occupancies as they do with the predictions of the arrangement of protons in the valence orbits.

The changes in nucleon occupancies during a potential double $\beta$ decay of $^{100}$Mo are also given in the Tables~\ref{ON-Table} and \ref{OP-Table} and displayed  graphically in Fig. \ref{CO}. For convenience, changes in the numbers of neutrons and protons are both quoted as positive numbers and therefore indicate the number of neutrons lost and the number of protons gained in the decay process. The neutron occupancy measurements indicate that the $1d$  (mainly the $j=5/2$ spin-orbit partner, assuming estimates above using existing assignments are correct) and  $0h_{11/2}$ orbits  participate strongly in a  double $\beta$ decay process between the ground states of the parent and daughter. There are smaller contributions from the $2s_{1/2}$ and $0g_{7/2}$ orbitals. The number of protons increases during the decay mainly in the $0g_{9/2}$ orbital, with the $1p$ protons (presumably with $j=1/2$) playing a lesser role and a much smaller contribution from the $0f_{5/2}$ orbital. 

Since the distribution of protons amongst the valence orbitals in the  parent and daughter nuclei are fairly well reproduced in the WS and IBM calculations, the picture of rearrangements of protons in such a decay are also reasonably well predicted overall, with some small differences in the contributions from different proton orbits as shown in Fig. \ref{CO}. The adjusted Woods-Saxon results appear to exaggerate the rearrangement of protons during a decay;  increases in $0g_{9/2}$ occupancy by more than two protons is compensated by {\it depletion} of proton $0f_{5/2}$ and $1p$ orbitals. Similarly, in the same calculation, more than two neutrons disappear from the $2s_{1/2}$ and $1d$ orbitals, balanced by {\it increases} in the $0h_{11/2}$ and $0g_{7/2}$ neutron occupancy. Such dramatic rearrangements are not substantiated in the experimental measurements for either type of nucleon. The predicted neutron occupancy changes in the WS and IBM calculations are rather similar to one another and to the experimental results for $2s_{1/2}$ and $1d$ neutrons, but the observed balance of neutron $0h_{11/2}$ and $0g_{7/2}$ contributions to the decay is not well reproduced.

None of the theoretical descriptions presented here reproduce all of the orbital occupancies and nucleon rearrangements, deduced from the current experimental work, that would occur during the double $\beta$ decay of $^{100}$Mo. The effect of the discrepancies on decay probability is somewhat difficult to judge without further theoretical investigation. Certainly the dramatic rearrangement of nucleons implicit in the adjusted Woods-Saxon calculations, which na\"ively might hinder a decay, seem unwarranted by the current results. These appear to arise mostly from problems with the adjustments made in the case of $^{100}$Mo.  Indeed, the data presented here and in the Supplemental Material \cite{Supplemental} for single-particle excitations in odd-A nuclei form a good basis on which to reassess the adjustments associated with both $^{100}$Mo and $^{100}$Ru, with additional constraining data for the other  nuclei populated in the current work on $^{98}$Mo and $^{102}$Ru targets. While the  IBM and unadjusted WS models seem to give a reasonable overall picture for protons, significant differences arise for the predicted neutron occupancies and rearrangements, particularly for the higher-$\ell$ orbits. It may prove instructive to determine the quantitative effect on the nuclear matrix element for $0\nu2\beta$ decay if these theoretical approaches were adjusted to more accurately reproduce the measured occupancies and also to extract theoretical occupancies at the QRPA level to refine the comparison with data presented here.

\section{conclusion}

We report on an experimental determination of neutron occupancies and proton vacancies from data on the ($d$,$p$), ($p$,$d$), ($^3$He,$\alpha$) and ($^3$He,$d$) reactions on  $^{98,100}$Mo and $^{100,102}$Ru isotopes. The work provides a detailed quantitative assessment of the rearrangements of protons and neutrons amongst the valence single-particle orbitals during double $\beta$ decay of $^{100}$Mo. There are significant disagreements with theoretical calculations of the same properties, calculations which have also been used to determine the nuclear matrix element for $0\nu2\beta$ decay. We hope that these data will stimulate further theoretical attention to refine future calculations of this quantity, which could be a critical component in  developing our understanding of the properties of neutrinos should the rare process of $0\nu2\beta$ decay ever be observed.
\\

\section{ACKNOWLEDGMENTS}

We would like to acknowledge the accelerator operating staff and target makers at the Maier-Leibnitz Laboratorium der M\"unchner Universit\"aten. This material is based upon work supported by the UK Science and Technology Facilities Council, the US Department of Energy, Office of
Nuclear Physics, under Contract No. DE-AC02-06CH11357,  the National Science Foundation Grant
No. PHY-08022648 (JINA) and the DFG Cluster of Excellence ``Origin and Structure of the Universe".


\end{document}


\title{Supplemental material for\\``An experimental study of the rearrangements of valence protons and neutrons amongst single-particle orbits during double {\boldmath$\beta$} decay in $^{100}$Mo.'' \\by S.J.~Freeman {\boldmath$et~al.$}}

\begin{abstract}
\vskip 3cm
The following material gives the state-by-state experimental data to supplement the main publication and should be read in conjunction with it.\\

The first set of data shown in Tables I to XVI give the absolute cross sections in units of mb/sr to final states populated in the reactions performed on  $^{98,100}$Mo  and $^{100,102}$Ru and targets. The errors quoted on the cross sections here are purely statistical and their absolute values are subject to a systematic error of the order of 5\%. The values of $\ell$ and $J^{\pi}$ quoted here are those used in the occupancy analysis as discussed in the main text. The $\ell$ values have generally been assigned on the basis of the angular distribution of cross section; where  assignments have been made previously, these assignments generally agreed with the literature. For neutron removal, the assignments have also considered the ratio between cross sections from ($p$,$d$) and ($^3$He,$\alpha$) reactions; this consideration concentrated on ratios at forward angles as discussed in the main text, unless explicitly noted here. A small number of states were assigned using this ratio only, again noted in the tables. The $J^{\pi}$ values quoted are either taken from the literature or are values consistent with the $\ell$ transfer; where literature gives a range of  $J^{\pi}$ values, in some circumstances the $\ell$ transfer inferred from the current data provides additional restrictions on spin.\\

The second set of data in Tables XVII to XXVIII give the spectroscopic factors in neutron addition and removal, and proton addition, for populated states normalised using the methodology described in the main publication and grouped according to $\ell$ transfer. Errors on the spectroscopic factors are dominated by the variation in the relative numbers due to the choice of potentials used in the DWBA calculations and are estimated to be up to 5\%.
\noindent 
\end{abstract}

\maketitle
\renewcommand{\thefootnote}{\alph{footnote}}
\newpage
\begingroup
\squeezetable
\begin{center}

\newcommand\T{\rule{0pt}{3ex}}
\newcommand \B{\rule[-1.2ex]{0pt}{0pt}}
\begin{longtable}{@{\extracolsep{\fill}}lccllllll@{}}
\caption{\label{tab1} Cross sections (mb/sr)  for states in $^{103}$Ru populated via the $^{102}$Ru($d$,$p$) reaction. Excitation energies are generally known to better than $\sim$3~keV. }\\
\hline \hline

$E$ (keV)\T\B & $\ell$& $J^{\pi}$ &{8$^{\circ}$} & 18$^{\circ}$  &27$^{\circ}$  &33$^{\circ}$ &43$^{\circ}$    \\

\hline\hline \endfirsthead

\\[-1.8ex] \hline
\multicolumn{8}{c}
{{ \tablename\ \thetable{} -- continued from previous page \T\B}} \\
\hline  
\endhead

\\[-1.8ex] \hline 
\multicolumn{8}{c}{{Continued on next page\ldots \T}} \\
\endfoot

\\[-1.8ex] \hline \hline
\endlastfoot

2.0$^{\footnotemark[1]}$&2&5/2$^+$&1.06(1)&4.43(1)&2.364(7)&1.098(4)&0.990(3)\\
134&2&5/2$^+$&0.006(1)&0.0236(6)&0.0131(5)&0.0092(3)&0.0085(3)\\
174&0&1/2$^+$&7.70(3)&0.975(4)&2.540(8)&2.468(7)&1.087(4)\\
213&4&7/2$^+$&0.098(3)&0.172(2)&0.262(2)&0.338(2)&0.246(1)\\
239&5&11/2$^-$&0.087(3)&0.228(2)&0.278(2)&0.363(2)&0.430(2)\\
297&3&7/2$^-$&0.292(5)&0.568(3)&0.536(3)&0.419(2)&0.237(1)\\
346&2&3/2$^+$&0.046(2)&0.165(2)&0.088(1)&0.0483(7)&0.0441(6)\\
406&2&3/2$^+$,   5/2$^+$&0.848(9)&2.687(8)&1.273(5)&0.646(3)&0.668(3)\\
\end{longtable}

\end{center}
\endgroup
\newpage
\begingroup
\squeezetable
\begin{center}
\newcommand\T{\rule{0pt}{3ex}}
\newcommand \B{\rule[-1.2ex]{0pt}{0pt}}
\begin{longtable}{@{\extracolsep{\fill}}lccllllll@{}}
\caption{\label{tab1} Cross sections (mb/sr)  for states in $^{101}$Ru populated via the $^{100}$Ru($d$,$p$) reaction. Excitation energies are generally known to better than $\sim$3~keV.  }\\
\hline \hline

$E$ (keV)\T\B & $\ell$& $J^{\pi}$ &{8$^{\circ}$} & 18$^{\circ}$  &27$^{\circ}$  &33$^{\circ}$ &43$^{\circ}$    \\

\hline\hline \endfirsthead

\\[-1.8ex] \hline
\multicolumn{8}{c}
{{ \tablename\ \thetable{} -- continued from previous page \T\B}} \\
\hline  
\endhead

\\[-1.8ex] \hline 
\multicolumn{8}{c}{{Continued on next page\ldots \T}} \\
\endfoot

\\[-1.8ex] \hline \hline
\endlastfoot

0&2&5/2$^+$&1.117(9)&3.93(1)&2.791(8)&1.155(4)&1.114(4)\\
128&2&3/2$^+$&0.035(2)&0.086(1)&0.0534(8)&0.0239(5)&0.0227(4)\\
308&4&7/2$^+$&0.187(4)&0.263(2)&0.318(3)&0.488(2)&0.396(2)\\
326&0&1/2$^+$&7.71(3)&0.734(3)&2.522(8)&2.375(6)&1.044(3)\\
421&2&3/2$^+$&0.070(3)&0.209(2)&0.139(1)&0.0631(8)&0.0653(7)\\
526&5&11/2$^-$&0.073(4)&0.134(2)&0.205(2)&0.277(2)&0.317(2)\\
534&2&3/2$^+$&0.415(7)&1.168(4)&0.683(4)&0.328(2)&0.371(2)\\
598&3&5/2$^-$&0.231(4)&0.336(2)&0.425(2)&0.328(2)&0.186(1)\\
\end{longtable}
\end{center}

\endgroup
\newpage

\begingroup
\squeezetable
\begin{center}
\newcommand\T{\rule{0pt}{3ex}}
\newcommand \B{\rule[-1.2ex]{0pt}{0pt}}
\begin{longtable}{@{\extracolsep{\fill}}lccllllll@{}}
\caption{\label{tab1} Cross sections (mb/sr)  for states in $^{101}$Mo populated via the $^{100}$Mo($d$,$p$) reaction.  Excitation energies are generally known to better than $\sim$3~keV.  }\\
\hline \hline

$E$ (keV)\T\B & $\ell$& $J^{\pi}$ &{8$^{\circ}$} & 18$^{\circ}$  &27$^{\circ}$  &33$^{\circ}$ &43$^{\circ}$    \\

\hline\hline \endfirsthead

\\[-1.8ex] \hline
\multicolumn{8}{c}
{{ \tablename\ \thetable{} -- continued from previous page \T\B}} \\
\hline  
\endhead

\\[-1.8ex] \hline 
\multicolumn{8}{c}{{Continued on next page\ldots \T}} \\
\endfoot

\\[-1.8ex] \hline \hline
\endlastfoot

0&0&1/2$^+$&7.54(3)&0.804(4)&2.492(9)&2.215(7)&1.072(4)\\
15&2&3/2$^{+}$&&0.089(1)&0.080(2)&0.047(1)&0.043(1)\\
57&2&5/2$^+$&1.05(1)&2.948(9)&1.956(7)&0.873(4)&0.771(3)\\
171&2&5/2$^+$&0.011(1)&0.0307(7)&0.0259(7)&0.0140(4)&0.0089(3)\\
237$^{\footnotemark[1]}$&2 \& 4&  3/2$^+$,   5/2$^+$ \& 7/2$^+$  &0.395(6)&0.557(3)&0.710(4)&0.648(3)&0.402(2)\\
272&5&11/2$^-$&0.062(3)&0.153(2)&0.255(3)&0.318(2)&0.340(3)\\
290&2&3/2$^+$&0.22(1)&0.593(7)&0.416(8)&0.185(6)&0.173(3)\\
294&0&1/2$^+$&0.87(1)&0.146(7)&0.233(9)&0.274(7)&0.132(3)\\
\end{longtable}
\end{center}

\endgroup
\newpage

\begingroup
\squeezetable
\begin{center}
\newcommand\T{\rule{0pt}{3ex}}
\newcommand \B{\rule[-1.2ex]{0pt}{0pt}}
\begin{longtable}{@{\extracolsep{\fill}}lccllllll@{}}
\caption{\label{tab1} Cross sections (mb/sr)  for states in $^{99}$Mo populated via the $^{98}$Mo($d$,$p$) reaction.  Excitation energies are generally known to better than $\sim$3~keV. }\\
\hline \hline

$E$ (keV)\T\B & $\ell$& $J^{\pi}$ &{8$^{\circ}$} & 18$^{\circ}$  &27$^{\circ}$  &33$^{\circ}$ &43$^{\circ}$    \\

\hline\hline \endfirsthead

\\[-1.8ex] \hline
\multicolumn{8}{c}
{{ \tablename\ \thetable{} -- continued from previous page \T\B}} \\
\hline  
\endhead

\\[-1.8ex] \hline 
\multicolumn{8}{c}{{Continued on next page\ldots \T}} \\
\endfoot

\\[-1.8ex] \hline \hline
\endlastfoot

0&0&1/2$^+$&9.56(5)&1.032(5)&3.26(1)&3.23(1)&1.365(5)\\
98&2&5/2$^+$&0.86(1)&2.91(1)&1.873(8)&0.918(5)&0.769(4)\\
235&4&7/2$^+$&0.196(7)&0.305(3)&0.560(4)&0.742(5)&0.443(3)\\
351&2&3/2$^+$&0.376(9)&0.903(5)&0.513(4)&0.263(3)&0.250(2)\\
525&0&1/2$^+$&0.92(1)&0.113(2)&0.312(3)&0.330(3)&0.144(2)\\
549&2&3/2$^+$&1.50(2)&3.72(1)&2.109(9)&1.156(6)&1.097(5)\\
615&2&5/2$^+$&0.084(5)&0.229(3)&0.154(2)&0.083(2)&0.068(1)\\
632&2&3/2$^+$&0.050(4)&0.081(2)&0.057(1)&0.028(1)&0.0287(7)\\
\end{longtable}
\end{center}

\endgroup
\newpage

\begingroup
\squeezetable
\begin{center}
\newcommand\T{\rule{0pt}{3ex}}
\newcommand \B{\rule[-1.2ex]{0pt}{0pt}}
\begin{longtable}{@{\extracolsep{\fill}}lccllll@{}}
\caption{\label{tab1} Cross sections (mb/sr)  for states in $^{101}$Ru populated via the $^{102}$Ru($p$,$d$) reaction.  Excitation energies are generally known to better than $\sim$2~keV. }\\
\hline \hline

$E$ (keV)\T\B & $\ell$& $J^{\pi}$ &{6$^{\circ}$} & 18$^{\circ}$  &31$^{\circ}$  &40$^{\circ}$    \\

\hline\hline \endfirsthead

\\[-1.8ex] \hline
\multicolumn{7}{c}
{{ \tablename\ \thetable{} -- continued from previous page \T\B}} \\
\hline  
\endhead

\\[-1.8ex] \hline 
\multicolumn{7}{c}{{Continued on next page\ldots \T}} \\
\endfoot

\\[-1.8ex] \hline \hline
\endlastfoot

0&2&5/2$^+$&1.275(7)&4.99(2)&1.172(6)&1.171(6)\\
127&(2)&3/2$^+$&0.0151(7)&0.0177(9)&0.0062(4)&0.0048(4)\\
307&4&7/2$^+$&0.150(6)&0.10(1)&0.29(1)&0.16(1)\\
311&2&5/2$^+$&0.065(5)&0.47(1)&0.15(1)&0.15(1)\\
326&0&1/2$^+$&3.10(1)&0.295(4)&0.547(4)&0.215(3)\\
422&2&3/2$^+$&0.134(2)&0.390(5)&0.077(2)&0.106(2)\\
527&5&11/2$^-$&0.037(3)&0.045(2)&0.113(3)&0.119(3)\\
534&2&3/2$^+$&0.179(3)&0.438(5)&0.085(2)&0.117(3)\\
\end{longtable}
\end{center}
\footnotetext[1]{A previously-known, but unresolved doublet of $\ell=2$ and $\ell=5$ transitions.}
\footnotetext[2]{An $\ell$ assignment has been made solely on the basis of the  ($^3$He,$\alpha$)/($p$,$d$)  cross-section ratio.}

\endgroup
\newpage

\begingroup
\squeezetable
\begin{center}
\newcommand\T{\rule{0pt}{3ex}}
\newcommand \B{\rule[-1.2ex]{0pt}{0pt}}
\begin{longtable}{@{\extracolsep{\fill}}lccllll@{}}
\caption{\label{tab1} Cross sections (mb/sr)  for states in $^{99}$Ru populated via the $^{100}$Ru($p$,$d$) reaction.  Excitation energies are generally known to better than $\sim$2~keV. }\\
\hline \hline
$E$ (keV)\T\B & $\ell$& $J^{\pi}$ &{6$^{\circ}$} & 18$^{\circ}$  &31$^{\circ}$  &40$^{\circ}$    \\
\hline\hline \endfirsthead
\\[-1.8ex] \hline
\multicolumn{7}{c}
{{ \tablename\ \thetable{} -- continued from previous page \T\B}} \\
\hline  
\endhead
\\[-1.8ex] \hline 
\multicolumn{7}{c}{{Continued on next page\ldots \T}} \\
\endfoot
\\[-1.8ex] \hline \hline
\endlastfoot
0&2&5/2$^+$&1.361(7)&5.40(1)&1.284(6)&1.246(8)\\
89&2&3/2$^+$&0.029(1)&0.084(2)&0.0174(9)&0.020(1)\\
322&2&3/2$^+$&0.162(3)&0.429(5)&0.083(2)&0.103(2)\\
341&4&7/2$^+$&0.097(2)&0.099(2)&0.257(3)&0.159(2)\\
443&0&1/2$^+$&1.602(7)&0.122(2)&0.261(2)&0.088(1)\\
575&2&5/2$^+$&0.0126(6)&0.058(2)&0.0130(6)&0.0147(7)\\
618$^{\footnotemark[1]}$&0 \& 4 &1/2$^+$ \& 7/2$^+$ &1.437(7)&0.161(2)&0.306(3)&0.101(2)\\
720&4&9/2$^+$&0.072(3)&0.054(1)&0.111(2)&0.141(3)\\
\end{longtable}
\end{center}
\footnotetext[1]{Doublet of previously known states: 617.89~keV 7/2$^+$ and 618.13~keV 1/2$^+$.}
\footnotetext[2]{An $\ell$ assignment on the basis of the  ($^3$He,$\alpha$)/($p$,$d$)  cross-section ratio at angles of 10$^{\circ}$ and 18$^{\circ}$ respectively.}
\footnotetext[3]{An $\ell$ assignment has been made solely on the basis of the  ($^3$He,$\alpha$)/($p$,$d$)  cross-section ratio.}

\endgroup

~~~~~

\newpage
\begingroup
\squeezetable
\begin{center}
\newcommand\T{\rule{0pt}{3ex}}
\newcommand \B{\rule[-1.2ex]{0pt}{0pt}}
\begin{longtable}{@{\extracolsep{\fill}}lccllll@{}}
\caption{\label{tab1} Cross sections (mb/sr)  for states in $^{99}$Mo populated via the $^{100}$Mo($p$,$d$) reaction. Excitation energies are generally known to better than $\sim$2~keV.}\\
\hline \hline

$E$ (keV)\T\B & $\ell$& $J^{\pi}$ &{6$^{\circ}$} & 18$^{\circ}$  &31$^{\circ}$  &40$^{\circ}$    \\

\hline\hline \endfirsthead

\\[-1.8ex] \hline
\multicolumn{7}{c}
{{ \tablename\ \thetable{} -- continued from previous page \T\B}} \\
\hline  
\endhead

\\[-1.8ex] \hline 
\multicolumn{7}{c}{{Continued on next page\ldots \T}} \\
\endfoot

\\[-1.8ex] \hline \hline
\endlastfoot
0&0&1/2$^+$&2.98(1)&0.307(4)&0.406(4)&0.248(2)\\
98&2&5/2$^+$&2.40(1)&7.61(2)&1.544(8)&1.634(7)\\
235&4&7/2$^+$&0.133(2)&0.156(3)&0.321(3)&0.150(2)\\
350&2&3/2$^+$&0.120(2)&0.350(4)&0.076(2)&0.074(1)\\
525&0&1/2$^+$&3.46(1)&0.327(4)&0.474(4)&0.248(4)\\
549&2&3/2$^+$&0.486(5)&1.177(8)&0.210(3)&0.287(4)\\
615&2&5/2$^+$&0.259(3)&0.962(7)&0.183(2)&0.215(3)\\
683&5&11/2$^-$&0.045(1)&0.064(2)&0.120(2)&0.134(2)\\
697&4&7/2$^+$&0.0227(9)&0.023(1)&0.042(1)&0.0245(9)\\
754&2& 3/2$^{+}$, 5/2$^{+}$ &0.046(1)&0.089(2)&0.059(1)&0.0375(9)\\
792&2&3/2$^+$&0.060(1)&0.226(3)&0.048(1)&0.049(1)\\
866&4&7/2$^+$$^{\footnotemark[1]}$&0.0159(7)&0.0188(9)&0.038(1)&0.0205(7)\\
891&2&3/2$^+$&0.070(2)&0.157(3)&0.032(1)&0.043(1)\\
906$^{\footnotemark[2]}$&0 \& 4 &1/2$^+$ \& 9/2$^{+}$ &0.943(4)&0.158(2)&0.266(2)&0.220(2)\\
945&2&5/2$^+$&0.136(2)&0.476(3)&0.103(1)&0.117(1)\\
1007&4&7/2$^+$,  9/2$^+$&0.0022(2)&0.0021(2)&0.0049(3)&0.0030(2)\\
1027&0&1/2$^+$&0.150(2)&0.046(1)&0.0360(7)&0.0155(5)\\
1048&4&7/2$^+$&0.0205(7)&0.0168(6)&0.0306(7)&0.0126(5)\\
1070&2&3/2$^+$,5/2$^+$&0.0067(4)&0.0141(6)&0.0032(3)&0.0038(3)\\
1143&4&7/2$^+$,  9/2$^+$&0.0089(4)&0.0090(4)&0.0188(5)&0.0136(4)\\
1167&2&5/2$^+$&0.088(1)&0.236(2)&0.0497(9)&0.067(1)\\
1197&2&3/2$^+$&0.040(1)&0.082(1)&0.0206(6)&0.0232(6)\\
1237&0&1/2$^+$&0.0109(4)&0.0010(2)&0.0016(2)&0.0006(1)\\
1272&(4)$^{\footnotemark[3]}$&     7/2$^+$,  9/2$^+$&& &			0.013(1)&	0.008(1)\\
1253&0&1/2$^+$&0.059(1)&0.0064(4)&0.0089(4)&0.0031(2)\\
1314&(4)$^{\footnotemark[4]}$&     7/2$^+$,  9/2$^+$&0.001(1)& &			0.004(1)&	0.003(1)\\
1342&4&7/2$^+$$^{\footnotemark[5]}$&0.0111(6)&0.008(1)&0.0226(6)&0.0210(6)\\
1367&(4)$^{\footnotemark[4]}$&     7/2$^+$,  9/2$^+$&0.002(1)	&		&	0.004(1)	&	0.004(1)\\
1444&2&3/2$^+$,  5/2$^+$&0.0081(6)&0.0122(6)&0.0040(3)&0.0046(4)\\
1465&4&9/2$^+$&0.0219(9)&0.0199(8)&0.040(1)&0.040(1)\\
1484&2&3/2$^+$,  5/2$^+$&0.051(1)&0.180(2)&0.036(1)&0.051(1)\\
1533&2&3/2$^+$,  5/2$^+$&0.075(2)&0.331(3)&0.070(1)&0.094(2)\\
1560&(0)&1/2$^+$&0.025(1)&0.004(1)&0.003(1)&\\
1572&2&3/2$^+$,  5/2$^+$&0.029(1)&0.071(2)&0.0104(5)&0.021(1)\\
1634&2&3/2$^+$,  5/2$^+$&0.0046(7)&0.0212(8)&0.0056(5)&0.0094(6)\\
1661&5&9/2$^-$,11/2$^-$&0.0065(5)&0.0154(7)&0.0222(9)&0.032(1)\\
1680&3&3/2$^-$,5/2$^-$&0.0049(5)&0.0074(5)&0.0082(6)&0.0050(5)\\
1726&2&3/2$^+$,  5/2$^+$&0.0209(9)&0.082(2)&0.0166(8)&0.022(1)\\
1739&2&3/2$^+$,  5/2$^+$&0.0204(9)&0.084(2)&0.0137(7)&0.0230(1)\\
1777&2&3/2$^+$,  5/2$^+$&0.0039(4)&0.0145(7)&0.008(1)&0.006(1)\\
1819&5&9/2$^-$,11/2$^-$&0.004(1)&&0.0210(8)&0.028(1)\\
1829&2&3/2$^+$,  5/2$^+$&0.0023(4)&0.028(1)&0.0041(4)&0.0062(6)\\
1855&(3)&3/2$^-$,5/2$^-$&0.0028(4)&0.0051(6)&0.0058(5)&0.0043(5)\\
1885&2&3/2$^+$,  5/2$^+$&0.0144(8)&0.042(1)&0.0063(5)&0.0104(9)\\
1916&0&1/2$^+$&0.174(3)&0.0143(7)&0.0172(8)&0.0072(7)\\
1930&1&1/2$^-$,  3/2$^-$&0.209(3)&0.172(3)&0.065(1)&0.047(2)\\
1950&0&1/2$^+$&0.110(2)&0.0211(8)&0.0205(8)&0.0063(9)\\
1973&2&3/2$^+$,  5/2$^+$&0.0143(7)&0.067(1)&0.0189(6)&0.018(1)\\
2005&1&1/2$^-$,  3/2$^-$&0.0180(6)&0.0230(6)&0.0067(3)&0.0083(5)\\
2038&(2)&3/2$^+$,  5/2$^+$&0.0127(5)&0.0220(6)&0.0141(4)&0.0132(7)\\
2057&2&3/2$^+$,  5/2$^+$&0.0136(6)&0.056(1)&0.0111(4)&0.0156(8)\\
2076&0&1/2$^+$&0.065(1)&0.0183(7)&0.0177(5)&0.0051(6)\\
2090&(4)$^{\footnotemark[4]}$&     7/2$^+$,  9/2$^+$&0.0035(5)&0.0124(6)&0.0088(4)&0.0076(8)\\
2125&4&9/2$^+$$^{\footnotemark[5]}$&0.08(1)&0.073(6)&0.135(2)&0.106(3)\\
2130&(1)&1/2$^-$,  3/2$^-$&0.15(1)&0.128(6)&0.1245(2)&0.110(2)\\
2171&(5)$^{\footnotemark[3]}$&   9/2$^-$,11/2$^-$  && &			0.006(1)&	0.002(1)\\
2183&1&1/2$^-$,  3/2$^-$&0.497(3)&0.422(3)&0.108(1)&0.117(2)\\
2210& (2)& 3/2$^+$,  5/2$^+$&0.009(1)	&0.015(1)	&0.002(1)	&0.004(1)\\
2233& & &0.011(1)&	0.004(1)&	0.003(1)\\
2260& (2)& 3/2$^+$,  5/2$^+$&0.005(1)&	0.018(1)&	0.006(1)\\
2298&(4/5)$^{\footnotemark[3]}$&     && &			0.005(1)&	0.004(1)\\
2345& 1& 1/2$^-$,  3/2$^-$&0.079(2)	&0.051(2)&	0.032(1)&	0.027(1)\\
2360& 1& 1/2$^-$,  3/2$^-$&0.054(2)&	0.041(1)&	0.024(1)&	0.020(1)\\
2373& 2&3/2$^+$,  5/2$^+$ &0.014(1)&	0.024(1)&	0.012(1)&	0.014(1)\\
2387& (2)& 3/2$^+$,  5/2$^+$&0.008(1)&	0.018(1)&	0.007(1)&	0.011(1)\\
2406& & &0.018(1)&	0.019(1)&	0.005(1)&	0.005(1)\\
2434&1&1/2$^-$,  3/2$^-$&0.127(3)&0.061(4)&0.026(1)&0.021(2)\\
2453 &2&3/2$^+$,  5/2$^+$& 0.017(1)&	0.028(1)&	0.007(1)&	0.007(1)\\
2466&1&1/2$^-$,  3/2$^-$&0.289(4)&0.319(4)&0.083(1)&0.096(3)\\
2533&1&1/2$^-$,  3/2$^-$&0.030(1)&0.023(1)&0.0064(6)&0.009(1)\\
2555&1&1/2$^-$,  3/2$^-$&0.136(3)&0.109(2)&0.0283(9)&0.041(2)\\
2582& & &0.008(1)&	0.011(1)&	0.016(1)&	0.005(1)\\
2564&1&1/2$^-$,  3/2$^-$&0.050(2)&0.043(2)&0.0110(7)&\\
2597&(1) & 1/2$^-$,  3/2$^-$&0.027(1)&	0.021(1)&	0.010(1)&	0.007(1)\\
2625&1&1/2$^-$,  3/2$^-$&0.136(2)&0.119(2)&0.0266(8)&0.034(2)\\
2641&0&1/2$^+$&0.058(2)&0.016(1)&0.0178(7)&0.009(1)\\
2664&1&1/2$^-$,  3/2$^-$&0.031(2)&0.063(2)&0.0083(6)&0.007(1)\\
2669&(1)&1/2$^-$,  3/2$^-$&0.062(3)&0.028(2)&0.0231(9)&0.026(2)\\
2708&1&1/2$^-$,  3/2$^-$&0.029(2)&0.039(2)&0.015(1)&0.010(1)\\
2720&1&1/2$^-$,  3/2$^-$&0.064(2)&0.080(3)&0.024(1)&0.026(2)\\
2739&1&1/2$^-$,  3/2$^-$&0.030(1)&0.055(2)&0.017(1)&0.015(1)\\
2759&1&1/2$^-$,  3/2$^-$&0.035(2)&0.051(4)&0.016(2)&0.019(1)\\
2767&1&1/2$^-$,  3/2$^-$&0.017(1)&0.015(3)&0.004(1)&\\
2801&1&1/2$^-$,  3/2$^-$&0.019(1)&0.028(2)&0.0118(7)&0.014(1)\\
2812&1&1/2$^-$,  3/2$^-$&0.030(1)&0.019(1)&0.0068(6)&0.0064(9)\\
2827&1&1/2$^-$,  3/2$^-$&0.034(1)&0.027(1)&0.0191(8)&0.016(1)\\
2876&1&1/2$^-$,  3/2$^-$&0.104(2)&0.110(2)&0.028(1)&0.036(2)\\
2897&1&1/2$^-$,  3/2$^-$&0.069(2)&0.095(2)&0.0246(9)&0.029(2)\\
2913& (1)& 1/2$^-$,  3/2$^-$&0.031(1)&	0.040(1)&	0.011(1)&	0.012(2)\\
2923& 2& 3/2$^+$,  5/2$^+$&0.005(1)&	0.017(1)&	0.004(1)&	0.007(2)\\
2943& & &0.016(1)&	0.022(1)&	0.010(1)&	0.008(1)\\
2961& (2)& 3/2$^+$,  5/2$^+$&0.004(1)&	0.018(1)&	0.003(1)&	0.005(1)\\
2975&(1)&1/2$^-$,  3/2$^-$&0.040	(1)&	0.040(1)&	0.009(1)&	0.015(1)\\
2987& & &0.024(1)&	0.010(1)&	0.013(1)&	0.015(1)\\
2999& (1)& 1/2$^-$,  3/2$^-$&0.065(2)&	0.075(2)&	0.031(1)&	0.030(1)\\
3019& & &0.033(1)&	0.046(1)&	0.037(1)&	0.041(2)\\
\end{longtable}
\end{center}
\footnotetext[1]{Spin assignment has been taken from Lhersonneau et al. Z.Phys A358, 317 (1997).}
\footnotetext[2]{A previously known, unresolved doublet of $\ell=0$ and $\ell=4$ transitions.}
\footnotetext[3]{An $\ell$ assignment from ($^3$He,$\alpha$)/($p$,$d$)  cross-section ratio at angles of 10$^{\circ}$ and 31$^{\circ}$.}
\footnotetext[4]{An $\ell$ assignment has been made solely on the basis of the  ($^3$He,$\alpha$)/($p$,$d$) cross-section ratio.}
\footnotetext[5]{Spin assignment has been taken from Hirowatari et al. Nucl. Phys. A 714 3 (2003). }

\endgroup
\newpage

\begingroup
\squeezetable
\begin{center}
\newcommand\T{\rule{0pt}{3ex}}
\newcommand \B{\rule[-1.2ex]{0pt}{0pt}}
\begin{longtable}{@{\extracolsep{\fill}}lccllll@{}}
\caption{\label{tab1} Cross sections (mb/sr)  for states in $^{97}$Mo populated via the $^{98}$Mo($p$,$d$)$^{97}$Mo reaction.  Excitation energies are generally known to better than $\sim$2~keV.}\\
\hline \hline

$E$ (keV)\T\B & $\ell$& $J^{\pi}$ &{6$^{\circ}$} & 18$^{\circ}$  &31$^{\circ}$  &40$^{\circ}$    \\

\hline\hline \endfirsthead

\\[-1.8ex] \hline
\multicolumn{7}{c}
{{ \tablename\ \thetable{} -- continued from previous page \T\B}} \\
\hline  
\endhead

\\[-1.8ex] \hline 
\multicolumn{7}{c}{{Continued on next page\ldots \T}} \\
\endfoot

\\[-1.8ex] \hline \hline
\endlastfoot
0&2&5/2$^+$&3.01(1)&10.77(3)&2.20(1)&2.29(3)\\
480&2&3/2$^+$&0.0128(9)&0.037(2)&0.0099(9)&0.0122(9)\\
658&4&7/2$^+$&0.067(2)&0.070(2)&0.145(3)&0.078(2)\\
679&0&1/2$^+$&2.99(1)&0.299(5)&0.447(6)&0.202(4)\\
720&2&5/2$^+$&0.20(4)&0.580(7)&0.104(3)&0.146(3)\\
888&0&1/2$^+$&0.429(6)&0.036(2)&0.055(2)&0.025(1)\\
1024&4&7/2$^+$$^{\footnotemark[1]}$&0.0150(8)&0.019(2)&0.0227(9)&0.0118(6)\\
1092&2&3/2$^+$&0.0073(6)&0.024(2)&0.0046(4)&0.0060(4)\\
1117&4&9/2$^+$&0.060(2)&0.050(4)&0.108(2)&0.121(2)\\
1265&2&3/2$^+$,  5/2$^+$&0.132(4)&0.30(5)&0.053(2)&0.084(5)\\
1270&2&5/2$^+$&0.043(3)&0.178(5)&0.045(2)&0.046(5)\\
1284&2&3/2$^+$,  5/2$^+$&0.119(2)&0.468(5)&0.096(2)&0.123(2)\\
1437&5&11/2$^-$&0.0101(8)&0.022(1)&0.038(1)&0.044(2)\\
1516&4&9/2$^+$&0.012(1)&0.016(1)&0.030(1)&0.032(1)\\
1548&0&1/2$^+$&0.183(4)&0.021(1)&0.031(1)&0.010(1)\\
1558&1&1/2$^-$,  3/2$^-$&0.023(2)&0.031(2)&0.009(1)&0.009(1)\\
1630&4&7/2$^{+}$&0.002(1)&0.003(1)&0.003(1)&0.002(1)\\
1698&4&9/2$^+$&0.0068(8)&0.0050(6)&0.0098(8)&0.0103(9)\\
1723&2&3/2$^+$,  5/2$^+$&0.0035(6)&0.016(1)&0.0038(6)&0.0053(8)\\
1733&(1)&1/2$^-$,  3/2$^-$&0.011(1)&0.007(1)&0.0040(6)&0.0016(6)\\
1787&(4)$^{\footnotemark[2]}$&7/2$^+$,  9/2$^+$&0.008(1)&0.0010(1)&0.010(1)&0.008(1)\\
1847&(1)&1/2$^-$,  3/2$^-$&0.0092(9)&0.0071(7)&0.0032(5)&0.0007(4)\\
1866&4&7/2$^+$,  9/2$^+$&0.0027(6)&0.0064(7)&0.0128(9)&0.014(1)\\
1959&(4)& 7/2$^+$  & &0.063(1)&0.043(1)&0.047(1)\\
2034&1&1/2$^-$,  3/2$^-$&0.081(2)&0.070(3)&0.016(1)&0.021(2)\\
2041&5&9/2$^-$,11/2$^-$&0.013(1)&0.030(2)&0.035(2)&0.047(2)\\
2089&5&9/2$^-$,11/2$^-$&0.0033(4)&&0.0093(5)&0.010(7)\\
2153&2&3/2$^+$,  5/2$^+$&0.051(1)&0.071(6)&0.0171(7)&0.025(1)\\
2176&0&1/2$^+$&0.138(2)&&0.0256(8)&0.0082(6)\\
2200&(4)$^{\footnotemark[3]}$&7/2$^+$,  9/2$^+$&&&0.006(1)&0.005(1)\\
2245&4&7/2$^+$,  9/2$^+$&0.0058(5)&0.0049(7)&0.010(5)&0.0080(6)\\
2267&1&1/2$^-$,  3/2$^-$&0.081(2)&0.071(2)&0.0214(7)&0.0195(9)\\
2315&4&7/2$^+$,  9/2$^+$&0.010(8)&0.0102(8)&0.021(1)&0.026(2)\\
2333&1&1/2$^-$,  3/2$^-$&0.037(2)&0.045(2)&0.0084(8)&0.009(1)\\
2385&1&1/2$^-$,  3/2$^-$&0.474(7)&0.495(7)&0.114(3)&0.134(3)\\
2412&2&5/2$^+$&0.009(1)&0.025(2)&&0.007(1)\\
2431&(1)&1/2$^-$,  3/2$^-$&0.0039(8)&0.0031(8)&&0.0028(8)\\
2452&(2)&3/2$^+$,  5/2$^+$&0.021(1)&0.031(1)&&0.009(1)\\
2463&1&1/2$^-$,  3/2$^-$&0.029(2)&0.017(1)&&0.0040(7)\\
2506&4&7/2$^+$,  9/2$^+$&0.042(3)&0.049(4)&0.083(7)&0.12(1)\\
2512&4&9/2$^+$&0.059(3)&0.071(4)&0.189(8)&0.13(1)\\
2553&(0)&1/2$^{+}$&0.009(1)&0.007(1)&0.006(1)\\
2620&(2)&3/2$^+$,  5/2$^+$&0.0018(6)&0.0093(9)&0.008(1)&0.0022(6)\\
2653&& &0.0086(9)&0.008(1)&0.0122(9)&\\
2677&(2)&3/2$^+$,  5/2$^+$&0.0054(8)&0.010(1)&0.0040(6)&\\
2706&1&1/2$^-$,  3/2$^-$&0.0048(8)&0.008(2)&0.0023(5)&\\
2745&1&1/2$^-$,  3/2$^-$&0.177(4)&0.184(4)&0.052(1)&0.058(3)\\
2776&(0)&1/2$^+$&0.062(2)&0.011(1)&0.0122(9)&\\
2791&1&1/2$^-$,  3/2$^-$&0.009(1)&0.012(1)&0.0028(6)&\\
2829&4&7/2$^+$,  9/2$^+$&0.017(1)&0.021(1)&0.021(1)&0.028(3)\\
2851&(1)&1/2$^-$,  3/2$^-$&0.022(1)&0.028(1)&0.0036(7)&0.015(2)\\
2875&4&7/2$^+$,  9/2$^+$&0.006(1)&0.011(1)&0.013(1)&0.013(3)\\
2908& & &0.004(1)&0.003(1)\\
2931&1&1/2$^-$,  3/2$^-$&0.0049(6)&0.0070(7)&0.0043(4)&0.0037(5)\\
2946&1&1/2$^-$,  3/2$^-$&0.0070(7)&0.0053(6)&0.0045(4)&0.0053(7)\\
2962&1&1/2$^-$,  3/2$^-$&0.014(1)&0.016(1)&0.0042(4)&0.0055(7)\\
2974&4&7/2$^+$,  9/2$^+$&0.0013(5)&0.0035(6)&0.0070(6)&0.0061(7)\\
2992&(1)&1/2$^-$,  3/2$^-$&0.0050(7)&0.0040(7)&0.0041(5)&0.0038(7)\\
3006&1&1/2$^-$,  3/2$^-$&0.125(3)&0.137(3)&0.044(1)&0.058(2)\\
3032&1&1/2$^-$,  3/2$^-$&0.0113(9)&0.010(1)&0.0042(6)&0.0044(9)\\
3046&1&1/2$^-$,  3/2$^-$&0.274(5)&0.259(5)&0.060(2)&0.069(2)\\
3075&1&1/2$^-$,  3/2$^-$&0.024(1)&0.029(2)&0.0051(6)&0.007(1)\\
3087& &&0.007(1)&0.003(1)\\
3098&&&0.004(1)&0.007(1)&0.008(1)&0.008(1)\\
3115&1&1/2$^-$,  3/2$^-$&0.129(3)&0.155(4)&0.032(1)&0.044(2)\\
3153&4&7/2$^+$,  9/2$^+$&0.013(1)&0.009(1)&0.020(1)&0.017(1)\\
3160&1&1/2$^-$,  3/2$^-$&0.027(2)&0.026(1)&0.0090(8)&0.013(1)\\
3171&4&7/2$^+$,  9/2$^+$&0.013(1)&0.015(1)&0.033(1)&0.038(2)\\
3183&1&1/2$^-$,  3/2$^-$&0.015(1)&0.019(1)&0.0062(8)&0.007(1)\\
3231& (4)$^{\footnotemark[2]}$&7/2$^+$,  9/2$^+$ &0.012(1)&0.008(1)\\
3240&&&0.005(1)&0.004(1)&0.011(1)&0.010(1)\\
3260&(1)&1/2$^-$,  3/2$^-$&0.075(2)&0.092(3)&0.0109(7)&0.0032(5)\\

\end{longtable}
\end{center}
\footnotetext[1]{At forward angles, there is possible small contribution from reactions on a target contaminant $^{95}$Mo leading to the population of the $\ell=2$ state at 2.3 MeV.}
\footnotetext[2]{An $\ell$ assignment has been made solely on the basis of the  ($^3$He,$\alpha$)/($p$,$d$)  cross-section ratio.}
\footnotetext[3]{An $\ell$ assignment has been  on the basis of the  ($^3$He,$\alpha$)/($p$,$d$) cross-section ratio at angles of 10$^{\circ}$ and 31$^{\circ}$.}

\endgroup
\newpage
\begingroup
\squeezetable
\begin{center}
\newcommand\T{\rule{0pt}{3ex}}
\newcommand \B{\rule[-1.2ex]{0pt}{0pt}}
\begin{longtable}{@{\extracolsep{\fill}}lccllll@{}}
\caption{\label{tab1} Cross sections (mb/sr)  for states in $^{101}$Ru populated via the $^{102}$Ru($^3$He,$\alpha$) reaction.  Excitation energies are known to better then $\sim$5~keV, rising to $\sim$8~keV above 2.7~MeV. }\\
\hline \hline

$E$ (keV)\T\B & $\ell$& $J^{\pi}$ &{10	$^{\circ}$} & 15$^{\circ}$  &20$^{\circ}$  &25$^{\circ}$    \\

\hline\hline \endfirsthead

\\[-1.8ex] \hline
\multicolumn{7}{c}
{{ \tablename\ \thetable{} -- continued from previous page \T\B}} \\
\hline  
\endhead

\\[-1.8ex] \hline 
\multicolumn{7}{c}{{Continued on next page\ldots \T}} \\
\endfoot

\\[-1.8ex] \hline \hline
\endlastfoot

0&2&5/2$^+$&0.510(8)&0.465(3)&0.357(2)&0.220(6)\\
129 & 2& 3/2$^+$& 0.005(1)&	0.004(1)&	0.003(1)&	0.003(1)\\
308&4&7/2$^+$&1.134(6)&0.670(3)&0.414(3)&0.262(3)\\
330&0&1/2$^+$&0.037(3)&0.018(1)&0.019(1)&0.016(2)\\
421&2&3/2$^+$&0.029(1)&0.0263(8)&0.0183(5)&0.0161(7)\\
527&5&11/2$^-$&1.22(5)&0.771(6)&0.421(5)&0.233(5)\\
548&4&7/2$^+$&0.17(1)&0.108(6)&0.074(6)&0.061(6)\\
598&3&5/2$^-$&0.052(9)&0.021(2)&0.015(1)&0.013(1)\\
614&2&3/2$^+$,  5/2$^+$&0.09(2)&0.077(2)&0.060(1)&0.036(2)\\
685&2&3/2$^+$,  5/2$^+$&0.026(1)&0.0214(8)&0.0095(7)&0.0068(6)\\
719&4&9/2$^+$&0.150(2)&0.094(1)&0.050(1)&0.043(1)\\
823&2&3/2$^+$,  5/2$^+$&0.028(6)&0.028(3)&0.023(2)&0.013(3)\\
841&4&7/2$^+$&0.103(6)&0.061(3)&0.032(2)&0.022(2)\\
930&4&9/2$^+$&0.116(4)&0.083(6)&0.047(4)&0.031(9)\\
1039&2&3/2$^+$,  5/2$^+$&0.050(2)&0.0442(9)&0.0305(7)&0.0223(8)\\
1153&4&7/2$^+$,  9/2$^+$&0.042(2)&0.0235(6)&0.0158(6)&0.0107(7)\\
1215&4&7/2$^+$,  9/2$^+$&0.053(2)&0.0354(9)&0.0198(7)&0.015(1)\\
1319 & (1)& 1/2$^-$,  3/2$^-$& 0.007(1)&	0.004(1)&	0.003(1)&	0.002(1)\\
1387&4&7/2$^+$,  9/2$^+$&0.023(1)&0.0164(7)&0.0084(4)&0.0076(6)\\
1426&4&7/2$^+$,  9/2$^+$&0.114(2)&0.078(1)&0.0456(8)&0.0311(9)\\
1478&2&3/2$^+$,  5/2$^+$&0.078(2)&0.063(1)&0.0391(7)&0.0287(9)\\
1538&2&3/2$^+$,  5/2$^+$&0.034(1)&0.0277(7)&0.0170(5)&0.0116(6)\\
1589&2&5/2$^+$&0.060(1)&0.0458(9)&0.0253(7)&0.0184(8)\\
1689$^{\footnotemark[1]}$&5& 9/2$^-$,11/2$^-$&	0.493(4)&	0.341(2)&	0.181(2)&	0.109(2)\\
1748$^{\footnotemark[2]}$ & (4/5)& & 0.012(1)&0.009(1)&0.005(1)\\
1819&2&3/2$^+$,  5/2$^+$& 0.043(1)&	0.035(1)&0.021(1)&	0.017(1)\\
1857 & 0&1/2$^{+}$ & 0.045(1)&0.030(1)&	0.015(1)&	0.010(1) \\
1891 & 4&7/2$^+$,  9/2$^+$ & 0.017(1)&0.011(1)&	0.006(1)&	0.005(1)\\
1940 & (4)$^{\footnotemark[3]}$&7/2$^+$,  9/2$^+$ & 0.048(2)&0.025(1)&	0.016(1)&0.011(1)\\
1970&2&3/2$^+$&0.060(2)&0.038(1)&	0.026(1)&	0.015(1)\\
1997&1&1/2$^-$,  3/2$^-$&0.053(2)&	0.045(1)&0.033(1)&0.025(2)\\
2067 & 2& 3/2$^+$,  5/2$^+$&0.020(2)&	0.016(1)&		0.007(1)&		0.006(1)\\
2131 & 1& 1/2$^-$,  3/2$^-$& 0.052(2)&	0.037(1)&		0.021(1)&		0.014(1)\\
2222&4&7/2$^+$,  9/2$^+$&0.102(3)&0.062(2)&0.032(2)&0.023(2)\\
2247&4&7/2$^+$,  9/2$^+$&0.093(3)&0.053(2)&0.025(1)&0.019(2)\\
2280&4&7/2$^+$,  9/2$^+$&0.442(5)&0.283(2)&0.150(2)&0.101(2)\\
2336&4&7/2$^+$,  9/2$^+$&0.685(5)&0.448(3)&0.241(2)&0.166(2)\\
2396&4&7/2$^+$,  9/2$^+$&0.723(5)&0.456(3)&0.238(2)&0.170(3)\\
2494&2&3/2$^+$,  5/2$^+$&0.091(3)&0.061(2)&0.035(1)&0.024(1)\\
2527&4&7/2$^+$,  9/2$^+$&0.165(4)&0.105(2)&0.060(2)&0.038(2)\\
2572&4&7/2$^+$,  9/2$^+$&0.113(4)&0.065(2)&0.038(2)&0.025(2)\\
2618&1&1/2$^-$,  3/2$^-$&0.052(3)&0.050(2)&0.034(2)&0.024(2)\\
2660&1&1/2$^-$,  3/2$^-$&0.021(3)&0.022(2)&0.017(2)&0.011(2)\\
2699&1&1/2$^-$,  3/2$^-$&0.039(4)&0.024(3)&0.012(2)&0.009(1)\\
2775&1&1/2$^-$,  3/2$^-$&0.035(3)&0.028(2)&0.017(1)&0.011(1)\\
2835&1&1/2$^-$,  3/2$^-$&0.055(6)&0.027(1)&0.024(1)&0.012(2)\\
2887&1&1/2$^-$,  3/2$^-$&0.034(4)&0.022(2)&0.013(2)&0.013(5)\\
2971&1&1/2$^-$,  3/2$^-$&0.027(2)&0.021(1)&0.0155(9)&0.009(1)\\
3033&1&1/2$^-$,  3/2$^-$&0.018(2)&0.018(1)&0.014(1)&0.0079(8)\\
3090&1&1/2$^-$,  3/2$^-$&0.044(3)&0.031(2)&0.020(1)&0.013(2)\\
3133&1&1/2$^-$,  3/2$^-$&0.047(3)&0.033(2)&0.021(1)&0.017(2)\\
3181&&&0.048(3)&0.039(2)&0.027(1)&0.019(2)\\
3238&&&0.055(3)&0.046(3)&0.023(2)&0.016(2)\\
3266&&&0.031(3)&0.021(2)&0.012(2)&0.013(2)\\
3302&&&0.020(2)&0.023(2)&0.012(1)&0.011(2)\\

\end{longtable}
\end{center}
\footnotetext[1]{There are potentially small contributions from a weak, but unresolved $\ell=2$ transition.}
\footnotetext[2]{There is no obvious counterpart for this state in the ($p$,$d$) reaction.}
\footnotetext[3]{An $\ell$ assignment has been made solely on the basis of the  ($^3$He,$\alpha$)/(p,d) cross-section ratio}

\endgroup
\newpage
\begingroup
\squeezetable
\begin{center}
\newcommand\T{\rule{0pt}{3ex}}
\newcommand \B{\rule[-1.2ex]{0pt}{0pt}}
\begin{longtable}{@{\extracolsep{\fill}}lccllll@{}}
\caption{\label{tab1} Cross sections (mb/sr)  for states in $^{99}$Ru populated via the $^{100}$Ru($^3$He,$\alpha$) reaction.  Excitation energies are known to better than $\sim$5~keV, rising to $\sim$10~keV above $\sim$2.8~MeV. }\\
\hline \hline

$E$ (keV)\T\B & $\ell$& $J^{\pi}$ &{10	$^{\circ}$} & 15$^{\circ}$  &20$^{\circ}$  &25$^{\circ}$    \\

\hline\hline \endfirsthead

\\[-1.8ex] \hline
\multicolumn{7}{c}
{{ \tablename\ \thetable{} -- continued from previous page \T\B}} \\
\hline  
\endhead

\\[-1.8ex] \hline 
\multicolumn{7}{c}{{Continued on next page\ldots \T}} \\
\endfoot

\\[-1.8ex] \hline \hline
\endlastfoot

0&2&5/2$^+$&0.651(4)&	0.577(3)&		0.440(3)&		0.268(3)	\\
89& 2&3/2$^+$ & 0.030(1)&	0.024(1)&	0.016(1)&		0.014(1)\\
340&4&7/2$^+$&1.017(6)&0.571(4)&0.358(3)&0.234(3)\\
440&0&1/2$^+$&0.007(2)&0.002(1)&0.004(1)&0.003(1)\\
575&2&5/2$^+$&0.013(1)&0.0096(6)&0.0068(4)&0.0048(4)\\
618$^{\footnotemark[1]}$& 4& 7/2$^{+}$&0.239(3)&0.118(1)&0.076(1)&0.053(1)\\
722&4&9/2$^+$&0.302(4)&0.180(2)&0.099(1)&0.077(2)\\
894&2&3/2$^+$,  5/2$^+$&0.034(1)&0.033(1)&0.0192(6)&0.0139(7)\\
998&2&3/2$^+$,  5/2$^+$&0.0290(9)&0.0310(7)&0.0149(5)&0.010(5)\\
1074&5&11/2$^-$&0.637(5)&0.431(3)&0.239(4)&0.162(3)\\
1198& & & 0.019(1)&	0.008(1)&	0.004(1)&	0.004(1)\\
1264&(4)$^{\footnotemark[2]}$&7/2$^+$&0.041(1)&	0.028(1)&		0.016(1)&		0.010(1)	\\
1304& 4& 7/2$^+$& 0.023(1)&	0.011(1)&		0.007(1)&		0.004(1)\\
1383& 0& 1/2$^+$& 0.005(1)&	0.002(1)&		0.003(1)&0.002(1)\\
1497&4&9/2$^+$&0.091(2)&0.062(1)&0.0339(9)&0.0245(9)\\
1583&4&7/2$^+$&0.075(2)&0.049(1)&0.0286(8)&0.0215(8)\\
1693&2&3/2$^+$,  5/2$^+$&0.074(2)&0.057(1)&0.0328(9)&0.0213(9)\\
1833&0&1/2$^+$&0.030(2)&0.021(1)&0.016(1)&0.013(1)\\
1943&5&11/2$^-$&0.322(4)&0.206(2)&0.110(2)&0.059(2)\\
2061& 1&   3/2$^-$& 0.025(1)&	0.018(1)&		0.008(1)&		0.007(1)\\
2119&4&7/2$^+$,  9/2$^+$&0.211(3)&0.127(2)&0.066(1)&0.042(2)\\
2168&4&7/2$^+$,  9/2$^+$&0.817(5)&0.527(4)&0.285(2)&0.179(3)\\
2232&4&7/2$^+$,  9/2$^+$&1.883(9)&1.173(7)&0.607(4)&0.397(4)\\
2350&(4)$^{\footnotemark[2]}$&7/2$^+$,  9/2$^+$&0.102(3)&0.064(2)&0.032(2)&0.023(2)\\
2404&0& 1/2$^+$&0.057(3)&0.033(1)&0.017(1)&0.011(1)\\
2493&1&1/2$^-$,  3/2$^-$&0.030(5)&0.028(2)&0.018(1)&0.013(1)\\
2540&1&1/2$^-$,  3/2$^-$&0.020(2)&0.033(1)&0.019(1)&0.012(1)\\
2599&1&1/2$^-$,  3/2$^-$&0.020(2)&0.020(1)&0.012(1)&0.0077(9)\\
2696&1&1/2$^-$,  3/2$^-$&0.078(2)&0.058(2)&0.044(2)&0.033(1)\\
2745&(4)$^{\footnotemark[3]}$&7/2$^+$,  9/2$^+$&0.096(2)&0.052(2)&0.026(2)&0.019(1)\\
2775&1&1/2$^-$,  3/2$^-$&0.049(2)&0.031(2)&0.019(2)&0.017(1)\\
2851& 1& 1/2$^-$,  3/2$^-$& 0.011(3)&		0.008(1)	\\
2895& 1& 1/2$^-$,  3/2$^-$& 0.013(3)&		0.008(1)	\\
2941& 1& 1/2$^-$,  3/2$^-$& 0.058(4)&		0.034(2)	&	0.029(3)&	0.015(2)\\
3014&1&1/2$^-$,  3/2$^-$&0.050(2)&0.031(1)&0.021(1)&0.017(1)\\
3068&1&1/2$^-$,  3/2$^-$&0.067(2)&0.042(1)&0.024(1)&0.014(1)\\
3110&1&1/2$^-$,  3/2$^-$&0.152(3)&0.105(2)&0.077(2)&0.058(2)\\

\end{longtable}
\end{center}
\footnotetext[1]{There are potentially small contributions from a weak, but unresolved $\ell=0$ transition.}
\footnotetext[2]{An $\ell$ assignment has been made solely on the basis of the  ($^3$He,$\alpha$)/(p,d) cross-section ratio.}
\footnotetext[3]{An $\ell$ assignment has been made  on the basis of the  ($^3$He,$\alpha$)/(p,d) cross-section ratio at angles of 10$^{\circ}$ and 18$^{\circ}$ respectively.}

\endgroup
\newpage

\begingroup
\squeezetable
\begin{center}
\newcommand\T{\rule{0pt}{3ex}}
\newcommand \B{\rule[-1.2ex]{0pt}{0pt}}
\begin{longtable}{@{\extracolsep{\fill}}lccllll@{}}
\caption{\label{tab1} Cross sections (mb/sr)  for states in $^{99}$Mo populated via the $^{100}$Mo($^3$He,$\alpha$) reaction.  Excitation energies are known to better than $\sim$5~keV, rising to $\sim$10~keV above $\sim$2.5~MeV. }\\
\hline \hline

$E$ (keV)\T\B & $\ell$& $J^{\pi}$ &{10	$^{\circ}$} & 15$^{\circ}$  &20$^{\circ}$  &25$^{\circ}$    \\

\hline\hline \endfirsthead

\\[-1.8ex] \hline
\multicolumn{7}{c}
{{ \tablename\ \thetable{} -- continued from previous page \T\B}} \\
\hline  
\endhead

\\[-1.8ex] \hline 
\multicolumn{7}{c}{{Continued on next page\ldots \T}} \\
\endfoot

\\[-1.8ex] \hline \hline
\endlastfoot
0&0&1/2$^+$&0.007(4)&0.002(1)&0.004(2)&0.005(2)\\
97&2&5/2$^+$&0.497(4)&0.458(3)&0.351(3)&0.217(3)\\
236&4&7/2$^+$&0.840(6)&0.444(3)&0.248(2)&0.162(2)\\
353&2&3/2$^+$&0.026(1)&0.0215(8)&0.0150(6)&0.0133(7)\\
524&0&1/2$^+$&0.008(2)& & 0.004(1)\\
549&2&3/2$^+$&0.061(2)&0.058(1)&0.045(1)&0.031(1)\\
615&2&5/2$^+$&0.086(2)&0.073(1)&0.057(1)&0.036(1)\\
686$^{\footnotemark[1]}$&4 \& 5&7/2$^+$ \& 11/2$^-$&1.180(7)&0.677(4)&0.385(3)&0.225(3)\\
752$^{\footnotemark[2]}$&2 \& 3& 3/2$^{+}$, 5/2$^{+}$ \& 7/2$^-$&0.038(2)&0.013(1)&0.0155(8)&0.009(1)\\
790&2&3/2$^+$&0.015(1)&0.012(1)&0.0119(8)&0.005(1)\\
867&4&7/2$^+$$^{\footnotemark[3]}$&0.114(2)&0.053(1)&0.037(1)&0.024(1)\\
905$^{\footnotemark[4]}$&0 \& 4&1/2$^+$ \& 9/2$^+$&0.267(3)&0.168(2)&0.090(2)&0.066(2)\\
944&2&5/2$^+$&0.039(2)&0.040(1)&0.0277(9)&0.0165(9)\\
1009&4&7/2$^+$,  9/2$^+$&0.016(1)&0.0074(5)&0.0045(4)&0.0035(7)\\
1049&4&7/2$^+$&0.152(2)&0.085(1)&0.049(1)&0.030(1)\\
1142&4&7/2$^+$,9/2$^+$&0.075(2)&0.045(2)&0.024(1)&0.017(2)\\
1166&2&5/2$^+$&0.057(2)&0.044(2)&0.030(1)&0.018(1)\\
1195&2&3/2$^+$&0.014(1)&0.013(1)&0.0074(7)&0.0047(8)\\
1273$^{\footnotemark[5]}$& (4)&7/2$^+$,9/2$^+$ & 0.032(1)&	0.019(1)&	0.011(1)&	0.007(1)\\
1313&4 & 7/2$^+$,9/2$^+$& 0.011(1)&0.007(1)&	0.005(1)&	0.003(1)\\
1342&4&7/2$^+$$^{\footnotemark[6]}$&0.059(2)&0.036(1)&0.022(1)&0.015(1)\\
1372&4&7/2$^+$,9/2$^+$&0.023(2)&0.015(2)&0.006(1)&0.004(1)\\
1468&4&9/2$^+$&0.137(3)&0.093(2)&0.054(1)&0.035(2)\\
1535&2&3/2$^+$,  5/2$^+$&0.102(3)&0.082(2)&0.050(2)&0.030(2)\\
1662&5&9/2$^-$,11/2$^-$&0.316(4)&0.209(3)&0.112(2)&0.058(2)\\
1732&2&3/2$^+$,  5/2$^+$&0.033(2)&0.026(2)&0.017(1)&0.008(1)\\
1818&5&9/2$^-$,11/2$^-$&0.257(5)&0.149(2)&0.079(2)&0.042(2)\\
1888&2&3/2$^+$,  5/2$^+$&0.020(1)&	0.013(1)&	0.006(1)&0.003(1)\\
1934&1&1/2$^-$,  3/2$^-$&0.027(3)&	0.017(2)&		0.014(1)&		0.012(1)\\
1966$^{\footnotemark[7]}$&(4/5) & &0.076(5)	&	0.050(2)	&	0.029(2)	&0.016(1)\\	
2038&2&3/2$^+$,  5/2$^+$&0.022(4)	&0.012(2)	&	0.005(1)	&	\\	
2087&(4)$^{\footnotemark[8]}$&7/2$^+$,  9/2$^+$&0.059(3)	&0.036(2)	&	0.019(1)	&0.008(1)	\\	
2127&4&9/2$^+$$^{\footnotemark[6]}$&1.124(7)&0.681(4)&0.370(3)&0.237(3)\\
2173$^{\footnotemark[5]}$& (5) & 9/2$^-$,11/2$^-$&0.086(3)&		0.060(2)&		0.041(1)&		0.026(1)\\	
2275&1&1/2$^-$,  3/2$^-$&0.059(8)&0.039(3)&0.017(3)&0.010(3)\\
2298$^{\footnotemark[5]}$&(4/5)&&0.025(6)&0.017(3)&0.007(1)&0.004(2)\\
2347&1&1/2$^-$,  3/2$^-$&0.047(2)&0.036(2)&0.023(2)&0.015(1)\\
2379&2&3/2$^+$,  5/2$^+$&0.039(2)&0.031(2)&0.017(1)&0.009(1)\\
2444&1&1/2$^-$,  3/2$^-$&0.019(2)&0.012(1)&0.012(1)&0.004(2)\\
2466&1&1/2$^-$,  3/2$^-$&0.045(2)&0.043(2)&0.022(1)&0.019(3)\\
2845$^{\footnotemark[7]}$&(4/5)&&0.078(4)&0.036(2)&0.018(1)&0.016(2)\\
2887&1&1/2$^-$,  3/2$^-$&0.070(3)&0.038(2)&0.020(1)&0.021(2)\\

\end{longtable}
\end{center}
\footnotetext[1]{The population of the 683-keV $\ell$=5 and  and 697-keV $\ell$=4 states, which were resolved in the ($p$,$d$) reaction, are unresolved in ($^3$He,$\alpha$) reaction. The spectroscopic strengths for  ($^3$He,$\alpha$) reaction were deduced on the basis of the relative spectroscopic strengths from the ($p$,$d$) reaction.}
\footnotetext[2]{A previously known, but unresolved doublet of $\ell=2$ and $\ell=3$ transitions.}
\footnotetext[3]{Spin assignment has been taken from Lhersonneau et al. Z.Phys A358, 317 (1997).}
\footnotetext[4]{There are potentially small contributions from a weak, but unresolved $\ell=0$ transition.}
\footnotetext[5]{An $\ell$ assignment has been made  on the basis of the  ($^3$He,$\alpha$)/(p,d) cross-section ratio at angles of 10$^{\circ}$ and 31$^{\circ}$.}
\footnotetext[6]{Spin assignment has been taken from Hirowatari et al. Nucl. Phys. A 714 3 (2003). }.
\footnotetext[7]{There is no obvious counterpart for this state in the ($p$,$d$) reaction.}
\footnotetext[8]{An $\ell$ assignment has been made solely on the basis of the  ($^3$He,$\alpha$)/($p$,$d$) cross-section ratio.}

\endgroup
\newpage
\begingroup
\squeezetable
\begin{center}
\newcommand\T{\rule{0pt}{3ex}}
\newcommand \B{\rule[-1.2ex]{0pt}{0pt}}
\begin{longtable}{@{\extracolsep{\fill}}lccllll@{}}
\caption{\label{tab1} Cross sections (mb/sr)  for states in $^{97}$Mo populated via the $^{98}$Mo($^3$He,$\alpha$)  reaction.  Excitation energies are known to better than $\sim$5~keV, rising to $\sim$10~keV above $\sim$3.0~MeV. }\\
\hline \hline

$E$ (keV)\T\B & $\ell$& $J^{\pi}$ &{10	$^{\circ}$} & 15$^{\circ}$  &20$^{\circ}$  &25$^{\circ}$    \\

\hline\hline \endfirsthead

\\[-1.8ex] \hline
\multicolumn{7}{c}
{{ \tablename\ \thetable{} -- continued from previous page \T\B}} \\
\hline  
\endhead

\\[-1.8ex] \hline 
\multicolumn{7}{c}{{Continued on next page\ldots \T}} \\
\endfoot

\\[-1.8ex] \hline \hline
\endlastfoot

0&2&5/2$^+$&0.756(6)&0.709(4)&0.557(3)&0.335(3)\\
478& 2& 3/2$^+$& 0.005(1)&	0.004(1)&		0.002(1)\\
659&4&7/2$^+$&0.463(4)&0.251(2)&0.155(2)&0.101(2)\\
682&0&1/2$^+$&0.061(2)&0.023(1)&0.018(1)&0.015(2)\\
719&2&5/2$^+$&0.062(2)&0.050(1)&0.035(1)&0.025(1)\\
887& 0&1/2$^+$ & 0.003(1)&		0.0010(5)&		0.0010	(5)&	\\
1028&4&7/2$^+$&0.107(2)&0.061(1)&0.0365(8)&0.022(1)\\
1118&4&9/2$^+$&0.216(3)&0.131(2)&0.072(1)&0.057(2)\\
1279$^{\footnotemark[1]}$&2&3/2$^+$,  5/2$^+$&0.152(3)&0.141(3)&0.087(1)&0.058(2)\\
1437&5&11/2$^-$&0.401(4)&0.262(2)&0.132(2)&0.080(2)\\
1514&4&9/2$^+$&0.075(3)&0.048(2)&0.028(1)&0.017(1)\\
1628&4&7/2$^+$&0.018(1)&0.0104(5)&0.0065(4)&0.0041(4)\\
1701&4&9/2$^+$&0.043(1)&0.0242(7)&0.0138(5)&0.0099(6)\\
1787$^{\footnotemark[4]}$&(4)&7/2$^+$, 9/2$^+$&0.030(1)&0.0172(6)&0.0103(4)&0.0070(5)\\
1865&4&7/2$^+$,  9/2$^+$&0.020(1)&0.0170(6)&0.0087(5)&0.0068(5)\\
1960&(4)&7/2$^+$&0.101(2)&0.071(1)&0.0388(8)&0.0257(9)\\
2043&5$^{\footnotemark[2]}$&9/2$^-$,11/2$^-$&0.666(5)&0.456(3)&0.242(2)&0.137(2)\\
2089&5&9/2$^-$,11/2$^-$&0.079(2)&0.047(1)&0.026(1)&0.013(1)\\
2155&2&3/2$^+$,  5/2$^+$&0.022(2)&0.023(1)&0.0130(7)&0.0069(6)\\
2198$^{\footnotemark[4]}$&(4) & 7/2$^+$, 9/2$^+$& 0.014(2)&	0.013(3)&	0.004(1)\\
2247&4&7/2$^+$,  9/2$^+$&0.034(2)&0.022(1)&0.0118(7)&0.0076(9)\\
2317&4&7/2$^+$,  9/2$^+$&0.065(2)&0.039(1)&0.0217(8)&0.0133(8)\\
2384&1&1/2$^-$,  3/2$^-$&0.057(2)&0.053(1)&0.037(1)&0.026(2)\\
2410&2&5/2$^+$&0.010(2)&0.0095(9)&0.0054(7)&0.005(2)\\
2510$^{\footnotemark[1]}$&4&7/2$^+$, 9/2$^+$\& 9/2$^+$&1.269(7)&0.784(4)&0.412(3)&0.266(3)\\
2830&4&7/2$^+$,  9/2$^+$&0.023(2)&0.0160(8)&0.0120(8)&0.0083(8)\\
2873&4&7/2$^+$,  9/2$^+$&0.037(2)&0.021(1)&0.0104(9)&0.009(1)\\
2935$^{\footnotemark[3]}$& (4/5)& & 0.020(3)&0.013(1)&0.010(1)\\
2974&4&7/2$^+$,  9/2$^+$&0.012(1)&0.014(2)&	0.011(1)&	0.004(1)\\
3009& 1&1/2$^-$,  3/2$^-$ & 0.061(2)&	0.049(1)&0.025(1)&0.020(2)\\
3049&1&1/2$^-$,  3/2$^-$&0.038	(2)&	0.038(1)&0.020(1)&0.019(1)\\
3110& 1&1/2$^-$,  3/2$^-$ & 0.030(2)&0.024(2)&0.014(1)&0.013(1)\\
3157&4&7/2$^+$,  9/2$^+$&0.073(8)&0.049(3)&0.018(7)&0.020(4)\\
3175&4&7/2$^+$,  9/2$^+$&0.20(9)&0.123(4)&0.065(8)&0.041(4)\\
3235&(4)$^{\footnotemark[2]}$&7/2$^+$,  9/2$^+$&0.040(2)&0.018(1)&0.015(2)&0.011(2)\\
3403&&(7/2$^+$,  9/2$^+$)$^{\footnotemark[5]}$&0.501(5)&0.30(3)&0.160(2)&0.104(2)\\
3573&&(7/2$^+$,  9/2$^+$)$^{\footnotemark[5]}$&0.114(4)&0.076(2)&0.039(2)&0.022(3)\\
3655&&(7/2$^+$,  9/2$^+$)$^{\footnotemark[5]}$&0.183(4)&0.127(3)&0.064(2)&0.039(2)\\

\end{longtable}
\end{center}
\footnotetext[1]{This unresolved multiplet of  $\ell=2$ transitions that is resolved in the ($p$,$d$) reaction.}
\footnotetext[2]{An $\ell$ assignment has been made solely on the basis of the  ($^3$He,$\alpha$)/(p,d) cross-section ratio.}
\footnotetext[3]{There is no obvious counterpart for this state in the ($p$,$d$) reaction.}
\footnotetext[4]{An $\ell$ assignment has been made on the basis of the  ($^3$He,$\alpha$)/(p,d) cross-section ratio at angles of 10$^{\circ}$ and 31$^{\circ}$.}
\footnotetext[5]{States assigned in a previous ($p$,$d$) measurement Bindal et al. Phys. Rev. C 12, 390 (1975), but lie beyond the excitation range measured in the current ($p$,$d$)  data.}

\endgroup
\newpage

\begingroup
\squeezetable
\begin{center}
\newcommand\T{\rule{0pt}{3ex}}
\newcommand \B{\rule[-1.2ex]{0pt}{0pt}}
\begin{longtable}{@{\extracolsep{\fill}}lcclllll@{}}
\caption{\label{tab1} Cross sections (mb/sr)  for states in $^{103}$Rh populated via the $^{102}$Ru($^3$He,$d$) reaction.  Excitation energies are generally known to better than $\sim$3~keV below $\sim$1.5~MeV, but rises to better an $\sim$10~keV at $\sim$3~MeV. }\\
\hline \hline

$E$ (keV)\T\B & $\ell$& $J^{\pi}$ &{6$^{\circ}$} & 10$^{\circ}$  &14$^{\circ}$  &18$^{\circ}$  &22$^{\circ}$  \\

\hline\hline \endfirsthead

\\[-1.8ex] \hline
\multicolumn{7}{c}
{{ \tablename\ \thetable{} -- continued from previous page \T\B}} \\
\hline  
\endhead

\\[-1.8ex] \hline 
\multicolumn{7}{c}{{Continued on next page\ldots \T}} \\
\endfoot

\\[-1.8ex] \hline \hline
\endlastfoot

0&1&1/2$^-$&1.789(7)&1.272(6)&0.434(2)&0.611(3)&0.470(3)\\
34&4&7/2$^+$&&&0.015(1)&0.022(1)&0.016(1)\\
92&4&9/2$^+$&0.121(2)&0.265(3)&0.499(3)&0.669(4)&0.475(3)\\
295&1&3/2$^-$&0.838(5)&0.630(4)&0.210(3)&0.274(2)&0.232(2)\\
357&3&5/2$^-$&0.009(2)&0.030(1)&0.0387(8)&0.0367(9)&0.0199(6)\\
649&2&5/2$^+$&0.595(4)&0.833(5)&0.519(3)&0.245(2)&0.299(2)\\
803&1&1/2$^-$&0.785(5)	&	0.586(4)&		0.206(2)&	0.264(2)&	0.224(2)\\
837&1&1/2$^-$&0.030(1)&	0.025(1)&	0.015(1)&	0.017(1)&	0.014(1)\\
\end{longtable}
\end{center}

\endgroup
\newpage

\begingroup
\squeezetable
\begin{center}
\newcommand\T{\rule{0pt}{3ex}}
\newcommand \B{\rule[-1.2ex]{0pt}{0pt}}
\begin{longtable}{@{\extracolsep{\fill}}lcclllll@{}}
\caption{\label{tab1} Cross sections (mb/sr)  for states in $^{101}$Rh populated via the $^{100}$Ru($^3$He,$d$) reaction. Excitation energies are generally known to better than $\sim$3~keV below $\sim$1.5~MeV, but rises to better an $\sim$10~keV at $\sim$3~MeV.}\\
\hline \hline

$E$ (keV)\T\B & $\ell$& $J^{\pi}$ &{6$^{\circ}$} & 10$^{\circ}$  &14$^{\circ}$  &18$^{\circ}$  &22$^{\circ}$  \\

\hline\hline \endfirsthead

\\[-1.8ex] \hline
\multicolumn{7}{c}
{{ \tablename\ \thetable{} -- continued from previous page \T\B}} \\
\hline  
\endhead

\\[-1.8ex] \hline 
\multicolumn{7}{c}{{Continued on next page\ldots \T}} \\
\endfoot

\\[-1.8ex] \hline \hline
\endlastfoot

0&1&1/2$^-$&1.767(7)&1.334(7)&0.459(3)&0.648(4)&0.533(3)\\
157&4&9/2$^+$&0.187(2)&0.374(4)&0.651(3)&0.903(4)&0.670(3)\\
306&1&3/2$^-$&0.655(4)&0.497(4)&0.159(2)&0.197(2)&0.177(2)\\
353&3&5/2$^-$&0.0128(9)&0.026(1)&0.0277(9)&0.017(1)&0.0177(9)\\
907&2&5/2$^+$&0.699(5)&0.942(6)&0.583(3)&0.296(3)&0.353(3)\\
998&1&3/2$^-$&0.780(4)&0.687(5)&0.231(2)&0.306(3)&0.267(2)\\
1058&3&5/2$^-$&0.038(5)&0.082(2)&0.098(2)&0.089(2)&0.042(1)\\
1307&4&7/2$^+$,  9/2$^+$&0.013(1)&0.020(1)&0.029(1)&0.034(1)&0.0232(9)\\
\end{longtable}
\end{center}

\endgroup
\newpage

\begingroup
\squeezetable
\begin{center}
\newcommand\T{\rule{0pt}{3ex}}
\newcommand \B{\rule[-1.2ex]{0pt}{0pt}}
\begin{longtable}{@{\extracolsep{\fill}}lcclllll@{}}
\caption{\label{tab1} Cross sections (mb/sr)  for states in $^{101}$Tc populated via the $^{100}$Mo($^3$He,$d$) reaction. Excitation energies are generally known to better than $\sim$3~keV. }\\
\hline \hline

$E$ (keV)\T\B & $\ell$& $J^{\pi}$ &{6$^{\circ}$} & 10$^{\circ}$  &14$^{\circ}$  &18$^{\circ}$  &22$^{\circ}$  \\
\hline\hline \endfirsthead

\\[-1.8ex] \hline
\multicolumn{7}{c}
{{ \tablename\ \thetable{} -- continued from previous page \T\B}} \\
\hline  
\endhead

\\[-1.8ex] \hline 
\multicolumn{7}{c}{{Continued on next page\ldots \T}} \\
\endfoot

\\[-1.8ex] \hline \hline
\endlastfoot

0&4&9/2$^+$&0.129(4)&0.283(5)&0.626(4)&0.812(4)&0.588(4)\\
209&1&1/2$^-$&1.852(8)&1.394(8)&0.420(3)&0.607(4)&0.434(3)\\
289&1&3/2$^-$&0.81(1)&0.726(6)&0.235(2)&0.326(3)&0.255(2)\\
393&3&5/2$^-$&&0.07(1)&0.073(1)&0.053(1)&0.0255(8)\\
514&2&5/2$^+$&0.441(4)&0.615(6)&0.346(3)&0.173(2)&0.211(2)\\
615&1&3/2$^-$&0.610(6)&0.433(5)&0.155(3)&0.169(3)&0.135(2)\\
674&3&5/2$^-$&0.033(2)&0.053(2)&0.054(2)&0.048(2)&0.027(1)\\
707& & &0..08(1)&0.006(1)& 0.002(1)& 0.004(1)& 0.002(1) \\
\end{longtable}
\end{center}

\endgroup
\newpage

\begingroup
\squeezetable
\begin{center}
\newcommand\T{\rule{0pt}{3ex}}
\newcommand \B{\rule[-1.2ex]{0pt}{0pt}}
\begin{longtable}{@{\extracolsep{\fill}}lcclll@{}}
\caption{\label{tab1} Cross sections (mb/sr)  for states in $^{99}$Tc populated via the $^{98}$Mo($^3$He,$d$) reaction. Excitation energies are generally known to better than $\sim$3~keV. }\\
\hline \hline

$E$ (keV)\T\B & $\ell$& $J^{\pi}$ &{6$^{\circ}$} & 14$^{\circ}$  &18$^{\circ}$    \\

\hline\hline \endfirsthead

\\[-1.8ex] \hline
\multicolumn{5}{c}
{{ \tablename\ \thetable{} -- continued from previous page \T\B}} \\
\hline  
\endhead

\\[-1.8ex] \hline 
\multicolumn{5}{c}{{Continued on next page\ldots \T}} \\
\endfoot

\\[-1.8ex] \hline \hline
\endlastfoot

0	&	4	&	9/2$^+$						&	0.195(2)	&	0.841(4)	&	1.081(5)	\\
144$^{\footnotemark[1]}$	&	1	&	1/2$^-$	&	1.218(6)	&	0.305(2)	&	0.423(3)	\\
180	&	2	&	5/2$^+$						&	0.112(2)	&	0.060(1)	&	0.041(1)	\\
509	&	1	&	3/2$^-$						&	0.955(5)	&	0.221(2)	&	0.292(3)	\\
622	&	4	&	9/2$^+$						&	0.024(1)	&	0.110(1)	&	0.134(2)	\\
669	&	1	&	3/2$^-$						&	0.083(2)	&	0.020(7)	&	0.029(1)	\\
719	&	4	&	7/2$^+$						&	0.020(1)	&	0.059(1)	&	0.044(1)	\\
762	&	2	&	5/2$^+$						&	0.434(3)	&	0.330(3)	&	0.172(2)	\\
\end{longtable}
\end{center}
\footnotetext[1]{Doublet of a $\ell=2$ transition with a weak $\ell=4$ $7/2^+$ contribution previously seen, but no evidence for the weak component has been observed here.}
\footnotetext[2]{Assignment of 9/2$^{+}$ was taken  from Crowe et al. PRC 57, 590 (1998).}
\endgroup
\newpage

\begingroup
\squeezetable
\begin{center}
\newcommand\T{\rule{0pt}{3ex}}
\newcommand \B{\rule[-1.2ex]{0pt}{0pt}}
\begin{longtable}{@{\extracolsep{\fill}}lccc@{}}
\caption{\label{tab1} Normalised spectroscopic strength for states associated with valence orbitals with $\ell=0$, 2, 4 and 5 in $^{101}$Ru populated by neutron-removal reactions. Strength associated with $\ell=0$ and 2 transitions is obtained from the ($p$,$d$) reaction and $\ell=4$ and 5 strength from the  ($^3$He,$\alpha$) reaction (see main text for details).}\\
\hline \hline

$E$ (keV)\T\B & $\ell$& $J^{\pi}$ &$C^2S$ \\

\hline \endfirsthead

\\[-1.8ex] \hline
\multicolumn{4}{c}
{{ \tablename\ \thetable{} -- continued from previous page \T\B}} \\
\hline  
\endhead

\\[-1.8ex] \hline \hline
\multicolumn{4}{c}{{Continued on next page\ldots \T}} \\
\endfoot

\\[-1.8ex] \hline\hline
\endlastfoot

0.326	&	0	&	1/2$^{+}$	&	0.153	\\
0.624	&	0	&	1/2$^{+}$	&	0.085	\\
1.099	&	0	&	1/2$^{+}$	&	0.002	\\
1.111	&	0	&	1/2$^{+}$	&	0.009	\\
1.714	&	0	&	1/2$^{+}$	&	0.017	\\
1.759	&	0	&	1/2$^{+}$	&	0.013	\\
1.859	&	0	&	1/2$^{+}$	&	0.001	\\
2.059	&	0	&	1/2$^{+}$	&	0.004	\\
\end{longtable}
\end{center}
\footnotetext[1]{States are assumed to have a 7/2$^{+}$ assignment (see main text for details).}
\endgroup
\newpage
\begingroup
\squeezetable
\begin{center}
\newcommand\T{\rule{0pt}{3ex}}
\newcommand \B{\rule[-1.2ex]{0pt}{0pt}}
\begin{longtable}{@{\extracolsep{\fill}}lccc@{}}
\caption{\label{tab1} Normalised spectroscopic strength for states associated with valence orbitals with $\ell=0$, 2, 4 and 5 in $^{99}$Ru populated by neutron-removal reactions. Strength associated with $\ell=0$ and 2 transitions is obtained from the ($p$,$d$) reaction and $\ell=4$ and 5 strength from the  ($^3$He,$\alpha$) reaction (see main text for details).}\\
\hline \hline

$E$ (keV)\T\B & $\ell$& $J^{\pi}$ &$C^2S$ \\

\hline \endfirsthead

\\[-1.8ex] \hline
\multicolumn{4}{c}
{{ \tablename\ \thetable{} -- continued from previous page \T\B}} \\
\hline  
\endhead

\\[-1.8ex] \hline \hline
\multicolumn{4}{c}{{Continued on next page\ldots \T}} \\
\endfoot

\\[-1.8ex] \hline\hline
\endlastfoot

0.443	&	0	&	1/2$^+$	&	0.083	\\
0.618	&	0	&	1/2$^+$	&	0.079	\\
1.383	&	0	&	1/2$^+$	&	0.031	\\
1.593	&	0	&	1/2$^+$	&	0.004	\\
1.840	&	0	&	1/2$^+$	&	0.020	\\
2.006	&	0	&	1/2$^+$	&	0.002	\\
2.337	&	0	&	1/2$^+$	&	0.006	\\
2.403	&	0	&	1/2$^+$	&	0.007	\\
\end{longtable}
\end{center}
\endgroup
\newpage
\begingroup
\squeezetable
\begin{center}
\newcommand\T{\rule{0pt}{3ex}}
\newcommand \B{\rule[-1.2ex]{0pt}{0pt}}
\begin{longtable}{@{\extracolsep{\fill}}lccc@{}}
\caption{\label{tab1} Normalised spectroscopic strength for states associated with valence orbitals with $\ell=0$, 2, 4 and 5 in $^{99}$Mo populated by neutron-removal reactions. Strength associated with $\ell=0$ and 2 transitions is obtained from the ($p$,$d$) reaction and $\ell=4$ and 5 strength from the  ($^3$He,$\alpha$) reaction (see main text for details).}\\
\hline \hline

$E$ (keV)\T\B & $\ell$& $J^{\pi}$ &$C^2S$ \\

\hline \endfirsthead

\\[-1.8ex] \hline
\multicolumn{4}{c}
{{ \tablename\ \thetable{} -- continued from previous page \T\B}} \\
\hline  
\endhead

\\[-1.8ex] \hline \hline
\multicolumn{4}{c}{{Continued on next page\ldots \T}} \\
\endfoot

\\[-1.8ex] \hline\hline
\endlastfoot

0.000	&	0	&	1/2$^+$	&	0.117	\\
0.525	&	0	&	1/2$^+$	&	0.142	\\
0.906	&	0	&	1/2$^+$	&	0.035	\\
1.027	&	0	&	1/2$^+$	&	0.007	\\
1.237	&	0	&	1/2$^+$	&	0.001	\\
1.253	&	0	&	1/2$^+$	&	0.003	\\
1.560	&	0	&	1/2$^+$	&	0.001	\\
1.916	&	0	&	1/2$^+$	&	0.009	\\
\end{longtable}
\end{center}
\footnotetext[1]{States are assumed to have a 7/2$^{+}$ assignment (see main text for details).}
\endgroup
\newpage
\begingroup
\squeezetable
\begin{center}
\newcommand\T{\rule{0pt}{3ex}}
\newcommand \B{\rule[-1.2ex]{0pt}{0pt}}
\begin{longtable}{@{\extracolsep{\fill}}lccc@{}}
\caption{\label{tab1} Normalised spectroscopic strength for states associated  valence orbitals with $\ell=0$, 2, 4 and 5 in $^{97}$Mo populated by neutron-removal reactions. Strength associated with $\ell=0$ and 2 transitions is obtained from the ($p$,$d$) reaction and $\ell=4$ and 5 strength from the  ($^3$He,$\alpha$) reaction (see text for details).}\\
\hline \hline

$E$ (keV)\T\B & $\ell$& $J^{\pi}$ &$C^2S$ \\

\hline \endfirsthead

\\[-1.8ex] \hline
\multicolumn{4}{c}
{{ \tablename\ \thetable{} -- continued from previous page \T\B}} \\
\hline  
\endhead

\\[-1.8ex] \hline \hline
\multicolumn{4}{c}{{Continued on next page\ldots \T}} \\
\endfoot

\\[-1.8ex] \hline\hline
\endlastfoot

0.679	&	0	&	1/2$^+$	&	0.124	\\
0.888	&	0	&	1/2$^+$	&	0.018	\\
1.548	&	0	&	1/2$^+$	&	0.009	\\
2.175	&	0	&	1/2$^+$	&	0.008	\\
2.553	&	0	&	1/2$^+$	&	0.008	\\
2.776	&	0	&	1/2$^+$	&	0.001	\\
0.000	&	2	&	5/2$^+$	&	2.772	\\
0.480	&	2	&	3/2$^+$	&	0.010	\\
\end{longtable}
\end{center}
\footnotetext[1]{States are assumed to have a 7/2$^{+}$ assignment (see main text for details).}
\endgroup
\newpage
\begingroup
\squeezetable
\begin{center}
\newcommand\T{\rule{0pt}{3ex}}
\newcommand \B{\rule[-1.2ex]{0pt}{0pt}}
\begin{longtable}{@{\extracolsep{\fill}}lccc@{}}
\caption{\label{tab1} Normalised spectroscopic strength for states associated with  $\ell=0$ and 2 valence orbitals in $^{103}$Ru populated by the ($d$,$p$) reaction (see main text for details).}\\
\hline \hline

$E$ (keV)\T\B & $\ell$& $J^{\pi}$ &$(2J+1)C^2S$ \\

\hline \endfirsthead

\\[-1.8ex] \hline
\multicolumn{4}{c}
{{ \tablename\ \thetable{} -- continued from previous page \T\B}} \\
\hline  
\endhead

\\[-1.8ex] \hline \hline
\multicolumn{4}{c}{{Continued on next page\ldots \T}} \\
\endfoot

\\[-1.8ex] \hline\hline
\endlastfoot

0.174	&	0	&	1/2$^+$	&	1.332	\\
0.432	&	0	&	1/2$^+$	&	0.044	\\
0.480	&	0	&	1/2$^+$	&	0.002	\\
0.554	&	0	&	1/2$^+$	&	0.059	\\
0.737	&	0	&	1/2$^+$	&	0.066	\\
1.107	&	0	&	1/2$^+$	&	0.100	\\
1.644	&	0	&	1/2$^+$	&	0.014	\\
1.729	&	0	&	1/2$^+$	&	0.008	\\
\end{longtable}
\end{center}
\endgroup
\newpage
\begingroup
\squeezetable
\begin{center}
\newcommand\T{\rule{0pt}{3ex}}
\newcommand \B{\rule[-1.2ex]{0pt}{0pt}}
\begin{longtable}{@{\extracolsep{\fill}}lccc@{}}
\caption{\label{tab1} Normalised spectroscopic strength for states associated with   $\ell=0$ and 2 valence orbitals in $^{101}$Ru populated by the ($d$,$p$) reaction (see main text for details).  }\\
\hline \hline

$E$ (keV)\T\B & $\ell$& $J^{\pi}$ &$(2J+1)C^2S$ \\

\hline \endfirsthead

\\[-1.8ex] \hline
\multicolumn{4}{c}
{{ \tablename\ \thetable{} -- continued from previous page \T\B}} \\
\hline  
\endhead

\\[-1.8ex] \hline \hline
\multicolumn{4}{c}{{Continued on next page\ldots \T}} \\
\endfoot

\\[-1.8ex] \hline\hline
\endlastfoot

0.326	&	0	&	1/2$^+$	&	1.220	\\
0.623	&	0	&	1/2$^+$	&	0.054	\\
1.098	&	0	&	1/2$^+$	&	0.028	\\
1.111	&	0	&	1/2$^+$	&	0.165	\\
1.2898	&	0	&	1/2$^+$	&	0.001	\\
1.4218	&	0	&	1/2$^+$	&	0.004	\\
1.713	&	0	&	1/2$^+$	&	0.007	\\
1.759	&	0	&	1/2$^+$	&	0.003	\\
\end{longtable}
\end{center}
\endgroup
\newpage
\begingroup
\squeezetable
\begin{center}
\newcommand\T{\rule{0pt}{3ex}}
\newcommand \B{\rule[-1.2ex]{0pt}{0pt}}
\begin{longtable}{@{\extracolsep{\fill}}lccc@{}}
\caption{\label{tab1} 
Normalised spectroscopic strength for states associated with   $\ell=0$ and 2 valence orbitals in $^{101}$Mo populated by the ($d$,$p$) reaction (see main text for details).  
}\\
\hline \hline

$E$ (keV)\T\B & $\ell$& $J^{\pi}$ &$(2J+1)C^2S$ \\

\hline \endfirsthead

\\[-1.8ex] \hline
\multicolumn{4}{c}
{{ \tablename\ \thetable{} -- continued from previous page \T\B}} \\
\hline  
\endhead

\\[-1.8ex] \hline \hline
\multicolumn{4}{c}{{Continued on next page\ldots \T}} \\
\endfoot

\\[-1.8ex] \hline\hline
\endlastfoot

0.000	&	0	&	1/2$^+$	&	1.219	\\
0.294	&	0	&	1/2$^+$	&	0.143	\\
0.541	&	0	&	1/2$^+$	&	0.119	\\
1.012	&	0	&	1/2$^+$	&	0.010	\\
1.035	&	0	&	1/2$^+$	&	0.005	\\
1.196	&	0	&	1/2$^+$	&	0.010	\\
1.737	&	0	&	1/2$^+$	&	0.066	\\
1.840	&	0	&	1/2$^+$	&	0.013	\\
\end{longtable}
\end{center}
\endgroup
\newpage
\begingroup
\squeezetable
\begin{center}
\newcommand\T{\rule{0pt}{3ex}}
\newcommand \B{\rule[-1.2ex]{0pt}{0pt}}
\begin{longtable}{@{\extracolsep{\fill}}lccc@{}}
\caption{\label{tab1} Normalised spectroscopic strength for states associated with   $\ell=0$ and 2 valence orbitals in $^{99}$Mo populated by the ($d$,$p$) reaction (see main text for details).  
}\\
\hline \hline

$E$ (keV)\T\B & $\ell$& $J^{\pi}$ &$(2J+1)C^2S$ \\

\hline \endfirsthead

\\[-1.8ex] \hline
\multicolumn{4}{c}
{{ \tablename\ \thetable{} -- continued from previous page \T\B}} \\
\hline  
\endhead

\\[-1.8ex] \hline \hline
\multicolumn{4}{c}{{Continued on next page\ldots \T}} \\
\endfoot

\\[-1.8ex] \hline\hline
\endlastfoot

0.000	&	0	&	1/2$^+$	&	1.398	\\
0.525	&	0	&	1/2$^+$	&	0.136	\\
0.906	&	0	&	1/2$^+$	&	0.056	\\
1.254	&	0	&	1/2$^+$	&	0.026	\\
1.916	&	0	&	1/2$^+$	&	0.068	\\
1.950	&	0	&	1/2$^+$	&	0.013	\\
2.079	&	0	&	1/2$^+$	&	0.029	\\
2.222	&	0	&	1/2$^+$	&	0.029	\\
\end{longtable}
\end{center}
\endgroup
\newpage

\begingroup
\squeezetable
\begin{center}
\newcommand\T{\rule{0pt}{3ex}}
\newcommand \B{\rule[-1.2ex]{0pt}{0pt}}
\begin{longtable}{@{\extracolsep{\fill}}lccc@{}}
\caption{\label{tab1} Normalised spectroscopic strength for states associated with valence orbitals in $^{103}$Rh  for $\ell=1$, 3 and 4 transitions is obtained from the ($^3$He,$d$) reaction (see main text for details).}\\
\hline \hline

$E$ (keV)\T\B & $\ell$& $J^{\pi}$ &$(2J+1)C^2S$ \\

\hline \endfirsthead

\\[-1.8ex] \hline
\multicolumn{4}{c}
{{ \tablename\ \thetable{} -- continued from previous page \T\B}} \\
\hline  
\endhead

\\[-1.8ex] \hline \hline
\multicolumn{4}{c}{{Continued on next page\ldots \T}} \\
\endfoot

\\[-1.8ex] \hline\hline
\endlastfoot

0.000	&	1	&	1/2$^-$&	0.756	\\
0.295	&	1	&	3/2$^-$	&	0.325	\\
0.804	&	1	&	1/2$^-$	&	0.310	\\
0.837	&	1	&	1/2$^-$,3/2$^-$	&	0.012	\\
1.493	&	1	&	1/2$^-$,3/2$^-$	&	0.017	\\
1.704	&	1	&	1/2$^-$,3/2$^-$	&	0.013	\\
0.357	&	3	&	5/2$^-$	&	0.133	\\
0.881	&	3	&	5/2$^-$	&	0.308	\\
1.665	&	3	&	5/2$^-$	&	0.463	\\
0.033	&	4	&	7/2$^+$	&	0.176	\\
0.092	&	4	&	9/2$^+$	&	3.627	\\
1.480	&	4	&	7/2$^+$,9/2$^+$&	0.357	\\

\end{longtable}
\end{center}
\endgroup

\begingroup
\squeezetable
\begin{center}
\newcommand\T{\rule{0pt}{3ex}}
\newcommand \B{\rule[-1.2ex]{0pt}{0pt}}
\begin{longtable}{@{\extracolsep{\fill}}lccc@{}}
\caption{\label{tab1} Normalised spectroscopic strength for states associated with valence orbitals in $^{101}$Rh for $\ell=1$, 3 and 4 transitions is obtained from the ($^3$He,$d$) reaction (see main text for details).}\\
\hline \hline

$E$ (keV)\T\B & $\ell$& $J^{\pi}$ &$(2J+1)C^2S$ \\

\hline \endfirsthead

\\[-1.8ex] \hline
\multicolumn{4}{c}
{{ \tablename\ \thetable{} -- continued from previous page \T\B}} \\
\hline  
\endhead

\\[-1.8ex] \hline \hline
\multicolumn{4}{c}{{Continued on next page\ldots \T}} \\
\endfoot

\\[-1.8ex] \hline\hline
\endlastfoot

0.000	&	1	&	1/2$^-$	&	0.685	\\
0.306	&	1	&	3/2$^-$	&	0.236	\\
0.998	&	1	&	3/2$^-$	&	0.268	\\
1.740	&	1	&	1/2$^-$,3/2$^-$	&	0.018	\\
0.353	&	3	&	5/2$^-$	&	0.084	\\
1.058	&	3	&	5/2$^-$	&	0.265	\\
0.157	&	4	&	9/2$^+$	&	4.305	\\
1.307	&	4	&	7/2$^+$,9/2$^+$	&	0.134	\\

\end{longtable}
\end{center}
\endgroup
\begingroup
\squeezetable
\begin{center}
\newcommand\T{\rule{0pt}{3ex}}
\newcommand \B{\rule[-1.2ex]{0pt}{0pt}}
\begin{longtable}{@{\extracolsep{\fill}}lccc@{}}
\caption{\label{tab1} Normalised spectroscopic strength for states associated with valence orbitals in $^{101}$Tc for $\ell=1$, 3 and 4 transitions is obtained from the ($^3$He,$d$) reaction (see main text for details).}\\
\hline \hline

$E$ (keV)\T\B & $\ell$& $J^{\pi}$ &$(2J+1)C^2S$ \\

\hline \endfirsthead

\\[-1.8ex] \hline
\multicolumn{4}{c}
{{ \tablename\ \thetable{} -- continued from previous page \T\B}} \\
\hline  
\endhead

\\[-1.8ex] \hline \hline
\multicolumn{4}{c}{{Continued on next page\ldots \T}} \\
\endfoot

\\[-1.8ex] \hline\hline
\endlastfoot

0.209	&	1	&	1/2$^-$	&	0.777	\\
0.289	&	1	&	3/2$^-$	&	0.319	\\
0.615	&	1	&	3/2$^-$	&	0.240	\\
1.049	&	1	&	1/2$^-$,3/2$^-$	&	0.014	\\
1.102	&	1	&	1/2$^-$,3/2$^-$	&	0.017	\\
1.188	&	1	&	1/2$^-$,3/2$^-$	&	0.084	\\
1.231	&	1	&	1/2$^-$,3/2$^-$&	0.013	\\
1.782	&	1	&	1/2$^-$,3/2$^-$	&	0.008	\\
2.024	&	1	&	1/2$^-$,3/2$^-$	&	0.014	\\
0.393	&	3	&	5/2$^-$	&	0.272	\\
0.674	&	3	&	5/2$^-$	&	0.194	\\
0.000	&	4	&	9/2$^+$	&	4.905	\\
0.777	&	4	&	7/2$^+$,9/2$^+$	&	0.063	\\
0.906	&	4	&	7/2$^+$,9/2$^+$&	0.183	\\
0.975	&	4	&	7/2$^+$,9/2$^+$	&	0.117	\\
1.250	&	4	&	7/2$^+$,9/2$^+$	&	0.101	\\
1.286	&	4	&	7/2$^+$,9/2$^+$	&	0.569	\\

\end{longtable}
\end{center}
\endgroup
\newpage
\begingroup
\squeezetable
\begin{center}
\newcommand\T{\rule{0pt}{3ex}}
\newcommand \B{\rule[-1.2ex]{0pt}{0pt}}
\begin{longtable}{@{\extracolsep{\fill}}lccc@{}}
\caption{\label{tab1} Normalised spectroscopic strength for states associated with valence orbitals in $^{99}$Tc for $\ell=1$, 3 and 4 transitions is obtained from the ($^3$He,$d$) reaction (see main text for details).}\\
\hline \hline

$E$ (keV)\T\B & $\ell$& $J^{\pi}$ &$(2J+1)C^2S$ \\

\hline \endfirsthead

\\[-1.8ex] \hline
\multicolumn{4}{c}
{{ \tablename\ \thetable{} -- continued from previous page \T\B}} \\
\hline  
\endhead

\\[-1.8ex] \hline \hline
\multicolumn{4}{c}{{Continued on next page\ldots \T}} \\
\endfoot

\\[-1.8ex] \hline\hline
\endlastfoot

0.144	&	1	&	1/2$^-$	&	0.458	\\
0.509	&	1	&	3/2$^-$	&	0.331	\\
0.669	&	1	&	3/2$^-$	&	0.028	\\
0.884	&	1	&	1/2$^-$,3/2$^-$	&	0.004	\\
1.321	&	1	&	3/2$^-$	&	0.037	\\
1.437	&	1	&	3/2$^-$	&	0.046	\\
0.002	&	4	&	9/2$^+$	&	5.594	\\
0.622	&	4	&	9/2$^+$	&	0.632	\\
0.719	&	4	&	7/2$^+$&	0.238	\\
1.202	&	4	&	9/2$^+$&	0.201	\\
1.425	&	4	&	9/2$^+$,(7/2$^+$)	&	0.354	\\

\end{longtable}
\end{center}
\endgroup

\newpage